\documentclass[onecolumn]{article}

\usepackage{geometry} 
\usepackage{bm}

\usepackage[titletoc,toc,page,header]{appendix} \usepackage{orcidlink}
\usepackage{authblk}
\usepackage{graphicx} \usepackage{amsmath} \usepackage{amssymb}
\usepackage{float}
\usepackage{multirow} \usepackage{tikz,pgf} \usepackage{cite}
\usepackage{amstext} \usepackage{subcaption} \usepackage{bm}
\usepackage{booktabs}
\usepackage{latexsym} \usepackage{color} \usepackage{slashed}
\usepackage{hyperref}
\usepackage{multicol}
\allowdisplaybreaks[4] \usetikzlibrary{trees}
\usetikzlibrary{decorations.pathmorphing}
\usetikzlibrary{decorations.markings} \usetikzlibrary{patterns}
\usetikzlibrary{quotes,angles} \usetikzlibrary{decorations}
\usepackage{mathrsfs} \usepackage{amsbsy} \usepackage[normalem]{ulem}
\usepackage{extarrows} \usepackage{widetext}
 
\begin{document}
\title{\Large\textbf{Lattice QCD study of the $K^*(892)$ resonance at
		the physical point }} \author{Qu-Zhi
	Li\(^{1, 2,\orcidlink{0009-0001-2640-1174}}\)~\footnote{liquzhi@swjtu.edu.cn},
	Chuan Liu\(^{2,3,4}\), Liuming Liu\(^{5,6}\)
	~\footnote{liuming@impcas.ac.cn}, Peng Sun\(^{5,6}\), Jia-Jun
	Wu\(^{6,7}\), Zhiguang Xiao\(^{1}\)~\footnote{xiaozg@scu.edu.cn}, Yi-Bo Yang\(^{7,\orcidlink{0000-0002-3213-0438}}\),
	Han-Qing Zheng\(^{1}\)~\footnote{zhenghq@scu.edu.cn} }
\affil{\(^1\)The School of Physical Science and Technology, Southwest Jiaotong University, Chengdu, Sichuan
	610065, P. R. China} 
\affil{\(^2\)College of Physics, Sichuan University, Chengdu, Sichuan
	610065, P. R. China} 
    \affil{\(^3\)School of Physics, Peking
	University, Beijing 100871, China} \affil{\(^4\)Center for High
	Energy Physics, Peking University, Beijing 100871, China}
\affil{\(^5\)Collaborative Innovation Center of Quantum Matter,
	Beijing 100871, China} \affil{\(^6\)Institute of Modern Physics,
	Chinese Academy of Sciences, Lanzhou, 730000, China}
\affil{\(^7\)School of Physical Sciences, University of Chinese
	Academy of Sciences, Beijing 100049, China} \affil{\(^8\)Southern
	Center for Nuclear-Science Theory (SCNT), Institute of Modern
	Physics, Chinese Academy of Sciences, Huizhou 516000, China}
\affil{\(^8\)Institute of Theoretical Physics, Chinese Academy of Sciences, Beijing 100190, China}
\maketitle

\begin{abstract}
	We present a lattice QCD study of the $K^*(892)$ resonance using
	eight $N_f=2+1$ Wilson-Clover ensembles with three lattice spacings
	and six pion masses ranging from 135 to 320 MeV. For each ensemble,
	a large number of finite volume energy levels in the $P$-wave $K\pi$
	channel are determined. The energy dependence of the scattering
	phase shift is then obtained from L\"uscher’s finite-volume
	method. To systematically assess parametrization dependence, the
	amplitude is described using three different models, which yield
	consistent results. The resulting phase shifts show a clear resonant
	behavior for all ensembles, and the corresponding $K^*(892)$
	resonance pole is identified on the second Riemann sheet in the
	complex energy plane. The pole positions are extrapolated to the
	physical pion mass and the continuum limit, yielding a $K^*(892)$
	resonance located at 
	{ 
	 $\sqrt{s_0} =  [882.10\binom{+7.36}{-10.29}(0.30) -i22.73(3.32)(0.63)] \mathrm{MeV}.$,
	 } 
	which is in excellent agreement with the experimental value. 
	{ 
	The quoted uncertainty includes the statistical uncertainty and systematic 
	uncertainties associated with the choice of fit windows and three parameterizations of the scattering amplitude.
	} 
	\end{abstract}

\section{Introduction}
Quantum Chromodynamics (QCD), as the fundamental theory of the strong
interaction, provides in principle a complete description of the
hadron spectrum and hadron-hadron interactions. However, its
confinement property at low energies makes first-principle studies of
hadron properties inherently non-perturbative and analytically
intractable. Lattice QCD addresses this challenge by formulating the
theory on a discrete spacetime lattice, enabling systematic numerical
computations from first principles. Although lattice QCD study of
ground state spectrum has achieved percent-level precision with fully
controlled systematics, precision studies of resonances-- requiring
the extraction of scattering amplitudes and pole positions -- remain
far less common and are still challenging.  A widely recognized method
to study scattering processes in lattice QCD is L\"uscher's
finite-volume method~\cite{Luscher:1986pf,Luscher:1990ux}, which
connects finite-volume energy levels and infinite-volume scattering
amplitudes. The original L\"uscher's formula for two identical
spinless particles in the rest frame was generalized to more complex
physical scenarios involving particles with spin and different masses,
boosted frames and coupled channels
~\cite{Rummukainen:1995vs,Fu:2011xz,Kim:2005gf,Bernard:2008ax,Leskovec:2012gb,Gockeler:2012yj,He:2005ey,Briceno:2014oea}.

As the simplest two-meson unequal-mass system with strangeness,
$\pi K$ scattering plays a pivotal role in understanding low-energy
strong interaction dynamics and spontaneous symmetry breaking in the
strange-meson sector. Experimental data of the $I=1/2$
$\pi K \to \pi K$ scattering channel showqs a slowly and monotonically
rising in the $S$-wave phase shift, while the $P$-wave phase shift
rises rapidly below the $K\eta$
threshold.
The $P$-wave behavior is well described by a Breit-Wigner form with a
single resonance corresponding to the $K^*(892)$. In contrast, the
broad $\kappa$ resonance, responsible for the rise in the $S$-wave
phase shifts, has been in controversy for a long time due to the
moderate phase-shift variation --- until it was firmly identified
through rigorous dispersive analyses~\cite{Zhou:2006wm,
	Pelaez:2020uiw}; see, e.g., Refs.~\cite{Yao:2020bxx, Pelaez:2020gnd}
for a comprehensive review.

Lattice studies of $\pi K$ scattering have extended from
determinations of the scattering length~\cite{Miao:2004gy, Beane_2006,
	Nagata:2008wk, Fu:2011wc, Lang:2012sv, Sasaki:2013vxa,
	Helmes:2018nug,Fu:2026ntz} to attempts to resolve the $\kappa$
resonance by searching for additional finite-volume energy
levels~\cite{Prelovsek:2010kg, Alexandrou:2012rm,
	Guo:2013nja}. Despite this development, only a few comprehensive
determinations of the energy-dependent $\pi K$ scattering amplitude
have been achieved. For the vector resonance $K^*(892)$, the first
lattice calculation was performed by Fu et al.~\cite{Fu:2012tj} using
staggered fermions, followed by a study with $N_f=2$ Wilson
quarks~\cite{Prelovsek:2013ela}. The RQCD
Collaboration~\cite{Bali:2015gji} repeated the calculation for
$K^*(892)$ with a pion mass closer to the physical value. Recently,
Boyle et al.~\cite{Boyle:2024hvv, Boyle:2024grr} presented the first
results for the $K^*(892)$ at physical quark masses.

The situation is more challenging for the $S$-wave $\kappa$
resonance. Not only are the $S$-wave and $P$-wave coupled in moving
frames within the L\"uscher formalism (ignoring higher partial waves
with $l \geq 2$), but the broad $\kappa$ resonance is also unstable
under simple $K$-matrix parameterizations. Rendon et
al.~\cite{Rendon:2020rtw} studied $S$- and $P$-wave $I=1/2$ $\pi K$
scatterings using $2+1$ flavors of dynamical clover fermions at
$m_\pi \simeq 317$ MeV and $176$ MeV. Their results indicate that the
narrow $P$-wave $K^*(892)$ resonance remains stable, while the broad
$\kappa$ resonance is only stable in the parameterizations that
include an Adler zero. A more comprehensive analysis by the Hadron
Spectrum Collaboration (HSC)~\cite{Dudek:2014qha, Wilson:2019wfr,
	Wilson:2014cna} suggests that at $m_\pi \simeq 391$ MeV, the
$S$-channel contains a virtual state associated with the $\kappa$,
while a clear shallow bound state corresponding to $K^*(892)$ appears
in the $P$-wave.

Resonances are rigorously identified as pole singularities of the
analytically continued scattering amplitude in the complex energy
plane. Broad resonances determined through model-dependent approaches,
such as $K$-matrix methods, can be problematic due to the difficulty
in quantifying systematic errors when extracting these pole
singularities. In contrast, the existence of broad
resonances can be reliably established using model-independent techniques like Roy or
Roy–Steiner equations~\cite{Roy:1971tc, Hite:1973pm}, which have
proven instrumental in determining the pole positions of particles
such as the $\sigma$, $\kappa$, and $N^*(920)$ in
$\pi\pi$~\cite{Caprini:2005zr}, $\pi K$~\cite{Descotes-Genon:2006sdr},
and $\pi N$~\cite{Ditsche:2012fv, Hoferichter:2015hva} systems,
respectively. These equations have also been employed to study
$\pi\pi$~\cite{Cao:2023ntr} and $\pi K$~\cite{Cao:2024zuy,
	Cao:2025hqm} scatterings at unphysical pion masses, utilizing
lattice QCD data~\cite{Briceno:2016mjc, Rodas:2023nec, Dudek:2014qha,
	Wilson:2014cna}. In doing so, intriguing phenomena have been
uncovered, such as a subthreshold resonance pole in the $\pi\pi$
system at $m_\pi \simeq 391$ MeV, findings that have been corroborated
by model studies~\cite{Lyu:2024lzr, Li:2025fvg}. Meanwhile,
Roy–Steiner analyses of the $\pi K$ system suggest that the $\kappa$
appears to remain as a broad resonance rather than evolving into a
virtual state as pion mass increases below
$m_\pi\simeq 391 $MeV~\cite{Cao:2024zuy,Cao:2025hqm}.

Besides the uncontrollable systematic errors, simple $K$-matrix
analyses of broad resonances also suffer from two main flaws. First,
The reliability of the $K$-matrix method is still
questionable~\cite{He:2002ut}, and thus whether the pole position is
trustworthy or not is not clear. In Roy–Steiner analyses, this domain
can be rigorously determined, allowing for confident identification of
resonance poles within it, even those far from threshold, such as the
$\sigma$ and $\kappa$ resonance. In contrast, such poles are less
trustworthy in a $K$-matrix approach, in particular, spurious poles on
the first Riemann sheet may also appears in some $K$-matrix
analysis. Second, the simple $K$-matrix approach neglects the
left-hand cut contributions to the phase shifts, which can be too
significant to ignore.

In this work, we present a systematic lattice study of $P$-wave
$I=1/2$ $\pi K$ scattering, based on eight ensembles of 2+1 flavor
Wilson-Clover gauge configurations covering three lattice spacings and
six different pion masses ranging from 135 to 320MeV. Using
L\"uscher's finite-volume method, we extract the mass and width of the
$K^*(892)$ resonance and extrapolate the results to the physical pion
mass and continuum limit. Our findings are consistent with the
experimental values. This work provides further insight into the
$K^*(892)$ resonance and its dependence on the pion mass, and
represents our first step toward a future lattice calculation of the
$S$-wave $\kappa$ resonance combining with chiral perturbation theory
and Roy-Steiner equations.

The paper is organized as follow: the lattice ensembles are listed in
Sec.~\ref{ensem}. Sec.~\ref{opera} and Sec.~\ref{corr} give details on
the constructions of interpolating operator and correlation
matrices. The determinations of finite-volume spectra are presented in
Sec.~\ref{sped}. Sec.~\ref{Lsa} show the strategies to fit the
finite-volume spectra via the L\"uscher formula. Three distinct
parameterizations for elastic $\pi K$ scattering in continuum space
are given in Sec.~\ref{PSC}.  Our numerical results and the
extrapolation are presented in
Sec.~\ref{numre}-\ref{extr}. Sec.~\ref{COt} gives a brief summary and
outlook. All fitted finite-volume spectra are listed in
App.~\ref{FVS}.

\section{Lattice methods}\label{ensem}
The results presented in this paper are based on the gauge
configurations generated by the CLQCD collaboration with 2+1 dynamical
quark flavors, employing the tadpole-improved Symanzik-improved gauge
action and Clover fermion action~\cite{CLQCD:2023sdb,CLQCD:2024yyn}.
We use eight ensembles with three lattice spacing values: $a$ =
0.105~fm, 0.077~fm and 0.052~fm, and various pion masses ranging from
135~MeV to 320~MeV.  Among these ensembles, F32P21/F48P21 and
F32P30/F48P30 are two pairs that share the same pion mass and lattice
spacing but different volumes to obtain more kinematic points in the
finite-volume spectra.  The ensemble C48P14 has the physical pion
mass.  The diverse parameters across these ensembles facilitate a
robust extrapolation of our results to the physical pion mass and
continuum limit.  The detailed information regarding these ensembles
is provided in Table~\ref{tab:conf}.

\begin{table}[htpb]
	\centering
	\begin{tabular}{|c|c|c|c|c|c|c|c|}
		\hline
		configuration & $L^3\times T$         & $a(\mathrm{fm})$
		              & $m_\pi(\mathrm{MeV})$ & $m_K(\mathrm{MeV})$ & $m_{\pi} L$ & $N_{\text{cfgs}}$ & $N_{ev}$             \\
		\hline
		C48P14        & $48^3\times 96$       & 0.10530(18)         & 134.2(1.6)  & 508.9(3.0)        & 3.56     & 261 & 150 \\
		\hline
		C48P23        & $48^3\times 96$       & 0.10530(18)         & 227.0(1.4)  & 483.7(2.9)        & 5.79     & 265 & 150 \\
		\hline
		C32P29        & $32^3\times 64$       & 0.10530(18)         & 293.6(1.8)  & 511.0(3.0)        & 5.01     & 984 & 100 \\
		\hline
		F32P21        & $32^3\times 64$       & 0.07746(18)         & 208.4(1.9)  & 491.8(1.3)        & 2.60
		              & 459                   & 100                                                                          \\
		\hline
		F48P21        & $48^3\times 96$       & 0.07746(18)         & 207.35(86)  & 491.3(1.2)        & 3.91     & 267 & 100 \\
		\hline
		F48P30        & $48^3\times 96$       & 0.07746(18)         & 304.91(81)  & 524.3(1.3)        & 5.72     & 359 & 100 \\
		\hline
		F32P30        & $32^3\times 96$       & 0.07746(18)         & 304.2(1.5)  & 524.5(1.5)        & 3.81     & 775 & 100 \\
		\hline
		H48P32        & $48^3\times 144$      & 0.05187(26)         & 318.4(1.7)  & 539.8(2.7)        & 4.06     & 453 & 100 \\
		\hline
	\end{tabular}
	\caption{Parameters of the ensembles and the propagators. The
		listed parameters are lattice size $(L/a)^3\times T/a$, the
		lattice spacing $a$, the mass of pion/kaon
		$m_{\pi}$/$m_{K}$, the value of $m_\pi L$, the number of
		configurations $N_{\mathrm{cfgs}}$ and the number of
		eigenvectors $N_{ev}$ used in the distillation method which
		will be introduced in the next section.}
	\label{tab:conf}
\end{table}

\section{Finite volume spectra}
\subsection{Interpolating operators}\label{opera}
The finite-volume spectrum is extracted from the correlation functions
of appropriate operators. The $SO(3)$ rotational symmetry of the
continuum space is broken down to the octahedral group $O_h$ on a
finite lattice and is further reduced to the little groups in the
moving frames. Consequently, the interpolating operators should
transform according to the irreducible representations (irreps) of the
$O_h$ group and its little groups.  In this work, we investigate
isospin-$\frac{1}{2}$ $P$-wave $K\pi$ scattering. In order to reliably
extract the complete low-energy spectrum, both quark bilinear
operators and $K\pi$ two-particle operators are needed. We consider
the operators in the rest frame as well as the moving frames with
total momenta $\boldsymbol P = (0,0,1)$, $(0,1,1)$, $(1,1,1)$ and
$(0,0,2)$, in units of $2\pi/L$.

For the isospin-$\frac{1}{2}$ $K\pi$ two-particle system with total
momentum $\boldsymbol P$ , the interpolating operators take the
general form:
\begin{equation}\label{DHO}
	O_{K \pi}^{\boldsymbol P, \Lambda, \lambda}\left(t, p_1, p_2 \right)  = \sum_{\boldsymbol{p}_1, \boldsymbol{p}_2 } c_{\Lambda, \lambda, \boldsymbol{p}_1, \boldsymbol{p}_2} \left[ \sqrt{\frac{2}{3}} \pi^{+}\left(t, \boldsymbol{p}_1\right) K^0\left(t, \boldsymbol{p}_2\right)
		-\sqrt{\frac{1}{3}} \pi^0\left(t, \boldsymbol{p}_1\right) K^{+}\left(t, \boldsymbol{p}_2\right)\right] , \quad \boldsymbol P = \boldsymbol p_1 + \boldsymbol p_2,
\end{equation}
where $\Lambda$ lables the irrep and $\lambda$ denotes its row. The
sum runs over all momenta $\boldsymbol{p}_1$($\boldsymbol{p}_2$) that
have the same magnitude $p_1(p_2)$ and are related by the rotations of
the octahedral group. The coefficients
$c_{\Lambda, \lambda, \boldsymbol{p}_1, \boldsymbol{p}_2}$ are chosen
so that the operator transforms under the irrep $\Lambda$. These
coefficients are constructed following the method of
Ref.~\cite{Yan:2025jlq} and are computed using the open-source package
\textbf{OpTion}~\cite{Yan:2025jlq} provided therein. The single particle operators for
$\pi$ and $K$ are the conventional quark bilinear operators. Besides
the $K\pi$ two-particle operators, we also include the single particle
operator for a vector meson:
\begin{equation}\label{SHO}
	O_{K^{*+}}^{\boldsymbol P, \Lambda, \lambda}(t) = \sum_{j} c_{\Lambda, \lambda, j} \sum_{\boldsymbol{x}} e^{i \boldsymbol{P} \cdot \boldsymbol{x}} \bar{s}(t, \boldsymbol{x}) \gamma_j u(t, \boldsymbol{x}), \quad j=1,2,3,
\end{equation}
where the coefficients $c_{\Lambda, \lambda, i}$ subduce a vector to
the lattice irreps and can also be computed using the package
\textbf{OpTion}.  Table~\ref{tab:LGIRs} lists all the frames and the
relevant irreps for which the P-wave is the lowest contributing
partial wave, together with the corresponding operators. Explicit
forms of all operators used in this work are provided in
Appendix~\ref{ops}.

\begin{table}[H]
	\centering
	\begin{tabular}{|c|c|c|c|c|}
		\hline
		$\boldsymbol{P}$           & LG($\boldsymbol P$)       & irrep $\Lambda$ & $SO(3)$                               & operators                                      \\
		\hline
		{$(0,0,0)$}                & {$O_h$}
		                           & $T_1^{-}$                 & $J=1,3\cdots$   & $K^{*+}_i,\pi_1K_1,\pi_2K_2,\pi_3K_3$                                                  \\
		\hline
		{$(0,0,1)$}                & {$C_{4v}$}                & $E$             & $J=1,2\cdots$                         & $K^{*+}_i,\pi_1K_2,\pi_2K_1,\pi_3K_2,\pi_2K_3$ \\
		\hline
		\multirow{2}{*}{$(0,1,1)$} & \multirow{2}{*}{$C_{2v}$}
		                           & $B_2$                     & $ J=1,2\cdots$  & $K^{*+}_i,\pi_2K_2,\pi_3K_1,\pi_1K_3$                                                  \\
		\cline{3-5}
		                           &                           & $B_1$           & $ J=1,2\cdots$                        & $K^{*+}_i,\pi_1K_1,\pi_2K_2,\pi_1K_5,\pi_2K_4$ \\
		\hline
		{$(1,1,1)$}                & {$C_{3v}$}                & $E$             & $J=1,2\cdots$                         & $K^{*+}_i,\pi_2K_1,\pi_1K_2,\pi_1K_6$          \\
		\hline
		{$(0,0,2)$}                & {$C_{4v}$}                & $E$             & $J=1,2\cdots$                         & $K^{*+}_i,\pi_2K_2,\pi_3K_3$                   \\
		\hline
	\end{tabular}
	\caption{The total momentum $\boldsymbol{P}$(in units of
		$2\pi/L$), the corresponding little groups and the irreps
		for which the P-wave is the lowest contributing partial
		wave. The last column presents the operators which would be
		projected into the corresponding irreps. The subscripts in
		the $K\pi$ operators are the square of the dimensionless
		momentum for $\pi$ and $K$ operators.  }
	\label{tab:LGIRs}
\end{table}

\subsection{Correlation matrix and Wick contractions}\label{corr}
The finite-volume spectrum in a specific irrep $\Lambda$ is obtained
by analyzing the correlation matrix $C^{\Lambda, \boldsymbol{ P}}(t)$
whose elements are the two-point correlation function of interpolators
$O^\Lambda_i$ defined above
\begin{equation}
	\begin{split}
		C_{ij}^{\Lambda,\boldsymbol{P}}(t)=\sum_{\lambda}\sum_{t_0=0}^T\langle O^{\boldsymbol{P},\Lambda,\lambda}_i(t + t_0, \boldsymbol P) O_j^{\boldsymbol{P},\Lambda,\lambda \,\dagger}(t_0,\boldsymbol P)\rangle_T~.
	\end{split}
\end{equation}
To improve statistical precision, the source time $t_0$ runs over all
time slices and all rows $\lambda$ are summed for multi-dimensional
irreps. The subscript $T$ indicates that the quantities performed on
the lattices are the thermal expectation values with the temporal lattice extent  $T$
-- the inverse of  temperature-- instead of vacuum
expectation values.

At finite $T$, the spectral decomposition of two-particle correlation
functions contains not only the desired signal
$\propto e^{-E_{K\pi}^n t}$(where $E_{K\pi}^n$ are the energies of the
$K\pi$ system) but also additional, unwanted contributions. For
example, the dominant extra term in the correlation function of the
operator $K(\boldsymbol{p_1})\pi( \boldsymbol{p_2})$ is proportional
to $e^{-TE_\pi(p_2)} e^{-t\left(E_K(p_1) - E_\pi(p_2)\right)}$,
assuming $E_\pi(p_2) < E_K(p_1)$. To remove the thermal states
contaminations, we follow the method proposed in
Ref.~\cite{Dudek:2012gj}. The correlation function is reweighted by a
factor and then shifted in the time direction, yielding the
weighted-shifted correlation function:
\begin{equation}
	\tilde{C} (t) = e^{t \Delta E} C(t) - e^{(t+\delta t) \Delta E} C(t+\delta t).
\end{equation}
where $\Delta E = E_K(p_1) - E_\pi(p_2)$ and should be the same for
all elements of the correlation matrix. The values of $p_1$ and $p_2$
are the momenta of the $K$ and $\pi$, respectively, of the $K\pi$
operator most affected by thermal pollution. The energies obtained
from the variational analysis(which will be introduced in the next
subsection) of the weighted-shifted correlation matrix are shifted
down by $\Delta E$, and it will restored to obtain the correct
energies. For simplicity, we use the same symbol
$C^{\Lambda,\boldsymbol P}(t)$ to represent the weighted-shifted
correlation matrix from here.

The elements of correlation matrix are expressed in terms of quark
propagators after Wick contractions.  The resulting quark-flow
diagrams are shown in Fig.~\ref{fig:quark-flow}.

\begin{figure}[H]
	\centering
	\begin{subfigure}[b]{0.25\textwidth}
		\centering
		\begin{tikzpicture}[node distance=1.5cm]
			\tikzstyle{strange}=[shape=circle, draw=black,
			fill=black!30] \tikzstyle{light}=[shape=circle,
			draw=black]
			\tikzstyle{itof}=[thick,decoration={markings,mark=at
					position 0.5 with
						{\arrow{>}}},postaction=decorate]
			\tikzstyle{ftoi}=[thick,decoration={markings,mark=at
					position 0.5 with
						{\arrow{<}}},postaction=decorate]

			\node[strange] at (0,0.5)
			(k_ti)[label=above:$t_2$,label=left:$\gamma_5$] {};
			\node[light] (pi_ti) [below
				of=k_ti,label=below:$t_2$,label=left:$\gamma_5$]
			{}; \node[strange] at (2,0.5)
			(k_tf)[label=above:$t_1$,label=right:$\gamma_5$]
			{} edge [ftoi, bend left=45]
			node[auto,swap]{$l$}(k_ti) edge [itof,bend
					right=45] node[auto,swap]{$s$}(k_ti);
			\node[light] (pi_tf) [below
				of=k_tf,label=below:$t_1$,label=right:$\gamma_5$]
			{} edge [ftoi, bend left=45]
			node[auto,swap]{$l$}(pi_ti) edge
				[itof,bend right=45]
			node[auto,swap]{$l$}(pi_ti);

		\end{tikzpicture}
		\caption*{A}
	\end{subfigure}
	\begin{subfigure}[b]{0.20\textwidth}
		\centering
		\begin{tikzpicture}[node distance=1.5cm]
			\tikzstyle{strange}=[shape=circle, draw=black,
			fill=black!30] \tikzstyle{light}=[shape=circle,
			draw=black]
			\tikzstyle{itof}=[thick,decoration={markings,mark=at
					position 0.5 with
						{\arrow{>}}},postaction=decorate]
			\tikzstyle{ftoi}=[thick,decoration={markings,mark=at
					position 0.5 with
						{\arrow{<}}},postaction=decorate]

			\node[strange] at (0,0.5)
			(k_tf)[label=above:$t_2$,label=left:$\gamma_5$] {};

			\node[light] (pi_tf) [below
				of=k_tf,label=below:$t_2$,label=left:$\gamma_5$]
			{};

			\node[strange] at (2,0.5)
			(k_ti)[label=above:$t_1$,label=right:$\gamma_5$]
			{} edge [itof] node[auto,swap]{$s$}(k_tf)
			edge [thick,decoration={markings,mark=at
							position 0.8 with
								{\arrow{<}}},postaction=decorate]
			node[auto,swap]{$l$}(pi_tf) ; \node[light]
			(pi_ti) [below
				of=k_ti,label=below:$t_1$,label=right:$\gamma_5$]
			{} edge
				[thick,decoration={markings,mark=at
							position 0.2 with
								{\arrow{<}}},postaction=decorate]
			node[auto,swap]{$l$}(k_tf) edge [itof]
			node[auto,swap]{$l$}(pi_tf);

		\end{tikzpicture}
		\caption*{X}
	\end{subfigure}
	\begin{subfigure}[b]{0.20\textwidth}
		\centering
		\begin{tikzpicture}[node distance=1.5cm]
			\tikzstyle{strange}=[shape=circle, draw=black,
			fill=black!30] \tikzstyle{light}=[shape=circle,
			draw=black]
			\tikzstyle{itof}=[thick,decoration={markings,mark=at
					position 0.5 with
						{\arrow{>}}},postaction=decorate]
			\tikzstyle{ftoi}=[thick,decoration={markings,mark=at
					position 0.5 with
						{\arrow{<}}},postaction=decorate]

			\node[strange] at (0,0.5)
			(k_ti)[label=above:$t_2$,label=left:$\gamma_5$] {};
			\node[light] (pi_ti) [below
				of=k_ti,label=below:$t_2$,label=left:$\gamma_5$]
			{} edge [ftoi] node[auto,swap]{$l$}(k_ti);
			\node[strange] at (2,0.5)
			(k_tf)[label=above:$t_1$,label=right:$\gamma_5$]
			{} edge [itof] node[auto,swap]{$s$}(k_ti);
			\node[light] (pi_tf) [below
				of=k_tf,label=below:$t_1$,label=right:$\gamma_5$]
			{} edge [itof]
			node[auto,swap]{$l$}(k_tf) edge
				[ftoi] node[auto,swap]{$l$}(pi_ti);
		\end{tikzpicture}
		\caption*{H}
	\end{subfigure}
	\begin{subfigure}[b]{0.20\textwidth}
		\centering
		\begin{tikzpicture}[node distance=1.5cm]
			\tikzstyle{strange}=[shape=circle, draw=black,
			fill=black!30] \tikzstyle{light}=[shape=circle,
			draw=black]
			\tikzstyle{itof}=[thick,decoration={markings,mark=at
					position 0.5 with
						{\arrow{>}}},postaction=decorate]
			\tikzstyle{ftoi}=[thick,decoration={markings,mark=at
					position 0.5 with
						{\arrow{<}}},postaction=decorate] \node[strange]
			at (0,0.5) (k_ti)
			[label=above:$t_1$,label=right:$\gamma_5$]{};
			\node[light] (pi_ti) [below of
				=k_ti,label=below:$t_1$,label=right:$\gamma_5$]{}
			edge [itof] node [auto,swap]{$l$} (k_ti);
			\node[strange] at (-1.2,-0.15) (k_tf) [
				label=below:$t_2$,label=left:$\gamma_i$]{} edge
				[itof] node [auto,swap]{$l$} (pi_ti) edge
				[ftoi] node[auto,swap]{$s$} (k_ti) ;
		\end{tikzpicture}
		\caption*{T}
	\end{subfigure}
	\caption{The Wick contractions corresponding to the elements
		$\langle (K\pi) (K\pi)^\dagger\rangle = A + \frac{1}{2} X -
			\frac{3}{2} H$ and
		$\langle K_i (K\pi)^\dagger \rangle = - \sqrt{ \frac{3}{2} }
			T$ for the correlation matrix
		$C^{\Lambda,\boldsymbol P}(t)$.}
	\label{fig:quark-flow}
\end{figure}

The quark propagators are computed using the distillation quark
smearing method~\cite{HadronSpectrum:2009krc}. The smearing operator
is defined as
\begin{equation}
	\Box(t) = V(t) V^\dagger(t),
\end{equation}
where $V(t)$ is a matrix of dimension $3L^3 \times N_{ev}$, whose
columns are the eigenvectors associated with the $N_{ev}$ lowest
eigenvalues of the three-dimensional Laplacian defined in terms of the
gauge fields:
\begin{align}\label{eq:Laplacian}
	-\nabla^2_{xy}=6\delta_{xy}-\sum^3_{j=1}\Big(\tilde{U}_j(x,t)\delta_{x+\hat{j},y}+\tilde{U}^{\dagger}_j(x-\hat{j},t)\delta_{x-\hat{j},y}\Big).
\end{align}
Here $\tilde{U}$ represents the gauge fields, to which HYP-smearing
has been applied in this work.

The smearing operator is applied on the quark fields both at sink and
source. After smearing, the quark propagator takes the form
\begin{equation}
	\mathcal{G}^{\boldsymbol{x}, t, \alpha, a}_{ \boldsymbol{x_0}, t_0, \alpha_0, a_0}  = V^{\boldsymbol{x}, t, a}_{i} (V^{\dagger})^i_{\boldsymbol{y}, t, b}  G^{\boldsymbol{y}, t, \alpha, b}_{\boldsymbol{y_0}, t_0, \alpha_0, b_0} V^{\boldsymbol{y_0}, t_0, b_0}_ j (V^{\dagger})^j_{\boldsymbol{x_0}, t_0, a_0},
\end{equation}
where $G$ stands for the exact all-to-all propagator. Explicit indices
for each matrix are given with the following notations:
$(\boldsymbol{x}$, $\boldsymbol{y}$, $\boldsymbol{x_0}$,
$\boldsymbol{y_0})$ label space positions, $(t, t_0)$ denote time
slices, $(\alpha, \alpha_0)$ represent Dirac indices, and
$(a, b, a_0, b_0)$ are color indices. Repeated indices are summed.

The exact all-to-all propagator $G$ is a matrix of dimension
$12 L^3 T \times 12 L^3 T$, which is prohibitively large to compute
and store directly.  However, the so called perambulator
\begin{equation}
	\mathcal{P}^{i, t, \alpha}_{j, t_0, \alpha_0} = (V^{\dagger})^i_{\boldsymbol{y}, t, b}  G^{\boldsymbol{y}, t, \alpha, b}_{\boldsymbol{y_0}, t_0, \alpha_0, b_0} V^{\boldsymbol{y_0}, t_0, b_0}_ j
\end{equation}
has the much smaller dimension $4 N_{ev} T \times 4 N_{ev} T$. Since
typically $N_{ev} \ll 3L^3$, the perambulator can be computed and
stored with affordable cost. The smeared quark propagator is then
constructed from the eigenvectors and perambulators:
$\mathcal{G} = V \mathcal{P} V^\dagger$.

The smeared quark propagator $\mathcal{G}$ is effectively an
all-to-all propagator. This property greatly facilitates the
calculation of quark-annihilation diagrams, such as the diagrams $H$
and $T$ in Fig.~\ref{fig:quark-flow}. It also enables the efficient
calculation of the correlation matrix involving many interpolating
operators and improve precision comparing to the conventional point-
and wall-source.  The configurations and perambulators used in this
work have been successfully utilized in many studies, see, e.g.,
~\cite{Yi:2025bnh,Shi:2025ogt, Yan:2025mdm,Wang:2025hew,Xing:2025uai,
	Yan:2024yuq,Yan:2024gwp,Liu:2026gxr} in hadron spectrum.

\subsection{Spectrum determination}\label{sped}
{ 
We determine the finite-volume spectrum by solving the generalized
eigenvalue problem (GEVP) for each correlation matrix
$C^{\Lambda,\boldsymbol{P}}$,
\begin{equation}
	C^ {\Lambda,\boldsymbol P }(t)\boldsymbol v_n = \lambda^{ \Lambda , \boldsymbol P}_n(t) C^ {\Lambda, \boldsymbol P}(t_0) \boldsymbol v_n~,
\end{equation}
where $t_0$ is the reference time. The values of $t_0$ used for each
irrep and ensemble are given in Appendix~\ref{FVS}. We extract the
energies $E_n^{\Lambda,\boldsymbol P}$ by fitting the principal
correlators $\lambda_n^{\Lambda,\boldsymbol P}(t)$ with the
two-exponential form
\begin{equation}
	\lambda^{ \Lambda,\boldsymbol P }_n(t) = (1-A_n)e^{-E^{\Lambda,\boldsymbol P}_n(t-t_0)}+A_n e^{-E^{\prime\Lambda,\boldsymbol P}_n(t-t_0)}~.
\end{equation}
The second exponential, with energy $E_n^{\prime\Lambda,\boldsymbol
P}$, accounts for residual excited-state contamination.

Statistical uncertainties are estimated using $N_{b} $ bootstrap
samples, where the first sample is the average over all configurations. In the following, we will index the first sample as $b=0$ and the rest of the samples as $b=1,\ldots, N_b-1$.  

To quantify the systematic uncertainty associated with the choice of
fit range, we consider every window $[t^l_{\min},t^l_{\max}]$ that is
contained in $[t_0+1,t_e]$ and satisfies the minimum-length condition
\begin{equation}
	[t^l_{\min},t^l_{\max}] \subseteq [t_0+1,t_e]~,
	\qquad t^l_{\max}-t^l_{\min} \geq 4~.
\end{equation}
The values of $t_e$ for all irreps and ensembles are listed in
Appendix~\ref{FVS}. They are chosen at a noisy-to-signal ratio cut of around $3\%$. This upper limit exclude very noisy late-time
data points from the fit. Such data points
contribute little to $\chi^2$ but increase the number of degrees of
freedom and can therefore artificially reduce $\chi^2/{\rm dof}$.

Following Refs.~\cite{Akaike:1974vps,Boyle:2024hvv,Boyle:2024grr},
we assign each fit window an Akaike-information-criterion (AIC) weight,
\begin{equation}
	\omega^l = \frac{1}{Z} e^{- \frac{1}{2} \text{AIC}^l}, \quad \text{AIC}^l= \chi_l^2  + 2 n ^{\text{para}} - n^l_{\text{data}}~.
	\label{eq:AIC}
\end{equation}
Here $n^{\text{para}}=3$ is the number of fit parameters and
$n^l_{\text{data}}$ is the number of data points in window $l$. The
normalization
$Z=\sum_l\exp(-\text{AIC}^l/2)$ ensures that $\sum_l\omega^l=1$.
The weights are determined from the $b=0$ data and then held fixed for
all bootstrap samples.

Let $E_{l,b}$ denote the energy obtained from fit window $l$ and
bootstrap sample $b$. For each bootstrap sample, we calculate the
AIC-weighted median $\widehat E_b$. Here we do not assume gaussian distribution over the fit windows, so we take the median instead of mean. 
We take $\widehat E_0$ as the
central value and the standard deviation of
$\{\widehat E_b\}_{b=0}^{N_b-1}$ as the statistical uncertainty $\sigma_{\mathrm{stat}}$. The
fit-window systematic uncertainty is asymmetric and the upper and lower uncertainties are estimated from the $b=0$ sample as: 
\begin{align}
	\sigma_{\mathrm{sys}}^+   = P_{84} - \widehat E_0, \quad \sigma_{\mathrm{sys}}^- = \widehat E_0 - P_{16} ,
	\label{eq:sys}
\end{align}
where $P_{16}$ and $P_{84}$ are the 16th and 84th
$\omega^l$-weighted percentiles of the distribution over fit windows.
The total uncertainty is
\begin{equation}\label{eq:totay}
\sigma_{\mathrm{tot}^{\pm}}=\sqrt{(\sigma_{\mathrm{stat}})^2+
(\sigma_{\mathrm{sys}}^{\pm})^2}.
\end{equation}

Figure~\ref{effe} shows representative effective masses and fit
results for the F48P30 ensemble, together with the corresponding
AIC-weighted distributions of energies obtained from all fit windows.
The bands indicate the total uncertainties defined above.
\begin{figure}[H]
	\centering
	\begin{subfigure}{0.48\textwidth}
		\includegraphics[width=\linewidth]{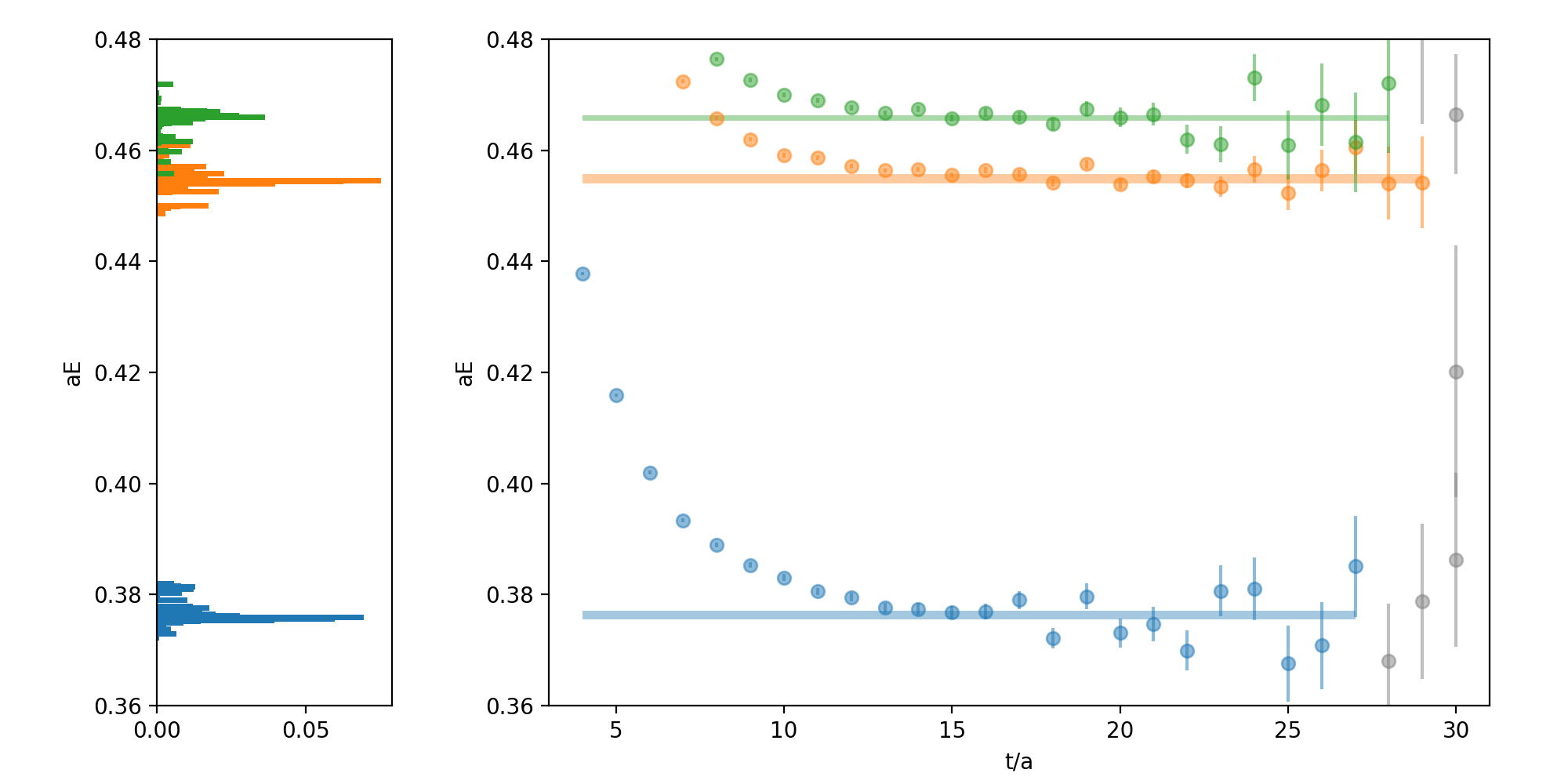}
		\caption{$(C_{4v},E)$}
	\end{subfigure}
	\hfill
	\begin{subfigure}{0.48\textwidth}
		\includegraphics[width=\linewidth]{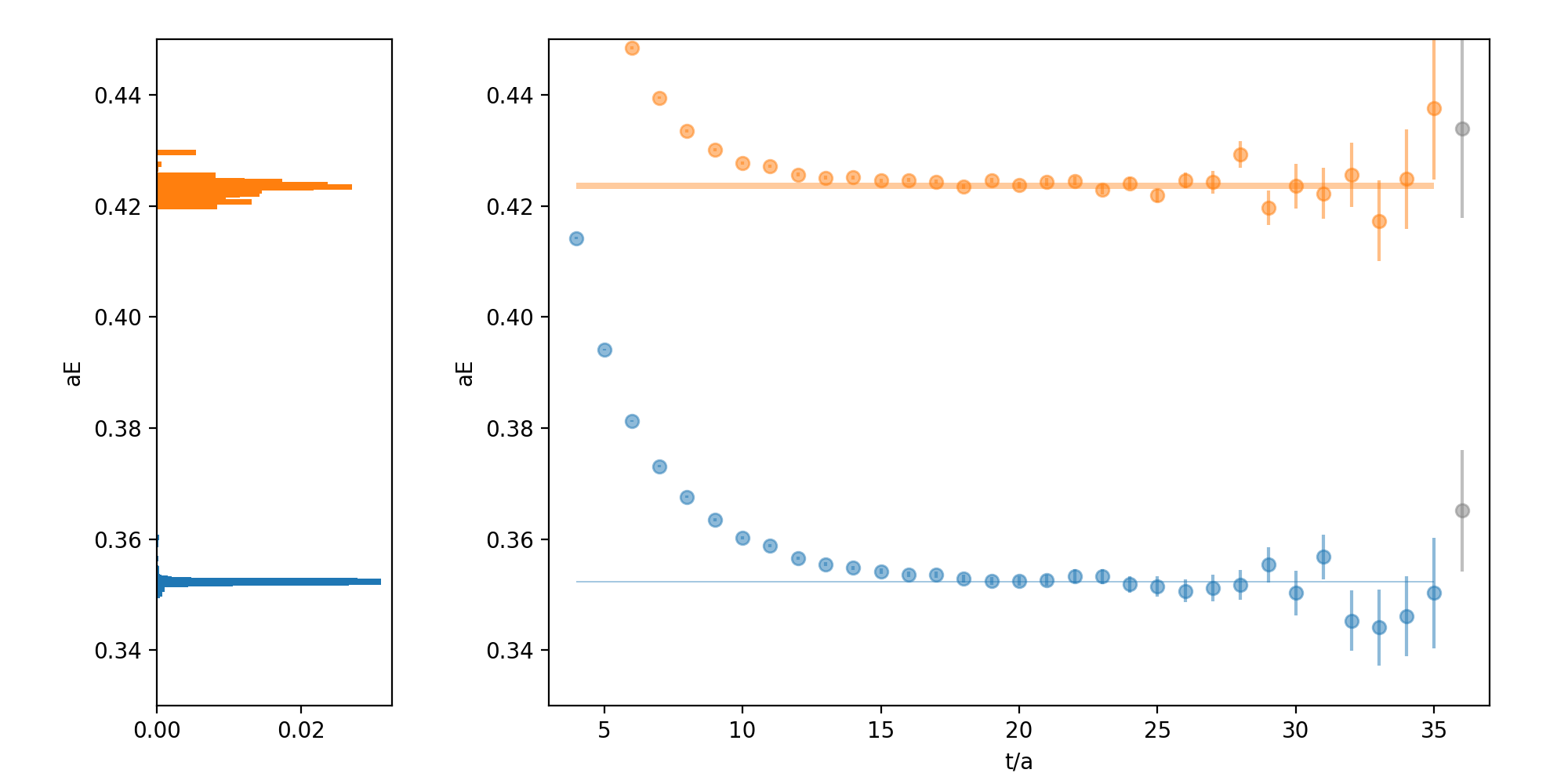}
		\caption{$(O_h,T_1^-)$}
	\end{subfigure}

	\caption{Effective masses of the GEVP principal correlators in the
		$(C_{4v},E)$ and $(O_h,T_1^-)$ irreps for the F48P30 ensemble.
		The histograms, shown in the corresponding colors, are the
		$\omega^l$-weighted distributions of energies obtained from
		two-exponential fits over all fit windows. The bands
		show the total uncertainties defined in the text.}\label{effe}
\end{figure}

The subsequent scattering analysis fits a set of
energy levels $\{E_1,E_2,\ldots,E_n\}$ using L\"uscher's formalism. If
level $E_i$ has $N_i$  fit windows, the total number of
possible fit-window combinations is
$N=\prod_{i=1}^n N_i$. Because an exhaustive analysis of all $N$
combinations is computationally prohibitive, we consider the following
two procedures for propagating both statistical and
fit-window systematics to the scattering parameters.
\begin{itemize}
	\item[a)] We draw $M$ fit-window combinations from the $N$
		possibilities, with $M\ll N$. A combination $f$ is sampled with
		weight
		$\omega_f=\prod_{i=1}^n\omega^{f(i)}$, where $f(i)$ labels the
		window selected for level $E_i$. We fit the L\"uscher's formula 
		for every combination $f$ and bootstrap sample $b$, obtaining
		scattering parameters $\alpha_b^f$. For each $b$, the weighted median over $f$ defines
		$\widehat\alpha_b$. The central value is
		$\widehat\alpha_0$, and the 16th and 84th percentiles of $\{\widehat\alpha_b\}_{b=0}^{N_b-1}$ gives the lower and upper statistical
		uncertainty. The fit-window systematic uncertainty is determined
		from the 16th and 84th weighted percentiles of
		$\{\alpha_0^f\}_{f=1}^M$.
	\item[b)] For each bootstrap sample $b$ and each energy level $E_i$,
		we select one fit window $l$ with probability $\omega^l$. Thus,
		each bootstrap sample is associated with one independently drawn
		combination of fit windows. After fitting the scattering
		parameters for every sample, we obtain
		$\{\alpha_b\}_{b=0}^{N_b-1}$. We quote the median of this
		distribution as the central value and its 16th and 84th
		percentiles as the uncertainty interval. In this procedure, the
		quoted uncertainty incorporates both statistical fluctuations and
		the systematic variation among fit windows.
\end{itemize}
Method a) requires $M\times N_{\mathrm{b}}$ fits to the L\"uscher
quantization condition. Repeating this procedure for eight ensembles
and three parameterizations of the scattering amplitude(see Sec.\ref{sec:ScatAna}) would be
computationally expensive. We therefore apply method a) only to the
C48P14 ensemble and use method b) for all ensembles. For method a), $N_b =2000$ and $M=300$. For method b), $N_b = 10000$. We compare the results from these two methods on ensemble C48P14 and found them consistent. The comparison is presented in Appendix~\ref{app:C48P14}. Our main results presented in Sec.~\ref{sec:ScatAna} are obtained from method b). 

Note that the energies sampled in method b) differ slightly
from the energies obtained using the genuine evaluation described
above. A comparison for all ensembles is presented in Appendix~\ref{app:compare}.

The finite-volume spectra for all ensembles are shown in
Figs.~\ref{fig:spectra1}--\ref{fig:spec-H48P32}. The energies shown
are those sampled using method b), because they serve as input to the
subsequent scattering analysis. Their medians are used as the central
values, while the 16th and 84th percentiles define the asymmetric
uncertainties. 
\begin{figure}[H]
	\centering \includegraphics[width=0.95\textwidth]{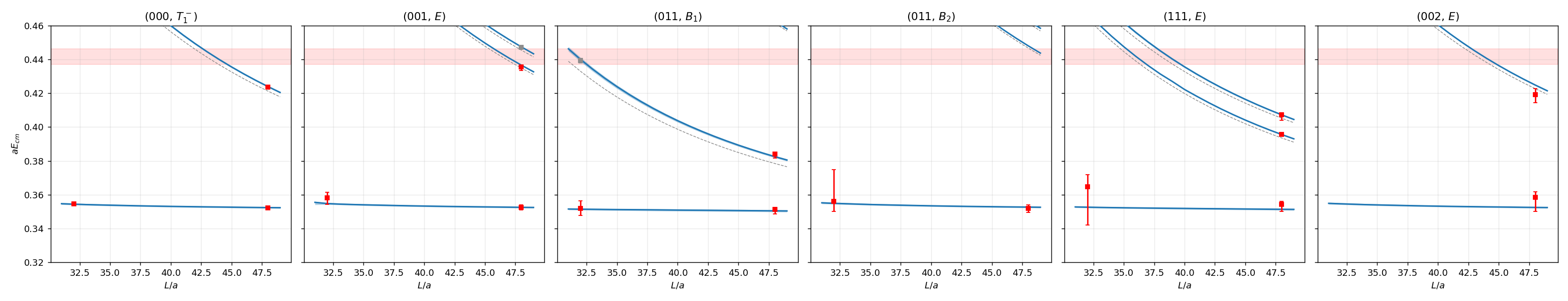}
	\caption{Finite-volume spectra, including total uncertainties, for
		the F48P30 and F32P30 ensembles. The red points are included in
		the scattering analysis described in Sec.~\ref{sec:ScatAna}.
		The gray dashed lines show the noninteracting energies, and the
		pink bands indicate the $K\eta$ threshold. The blue bands are
		the solutions of the L\"uscher quantization condition obtained
		with the PR parameterization; see Sec.~\ref{sec:ScatAna} for
		details.}
	\label{fig:spectra1}
\end{figure}
\begin{figure}[H]
	\centering \includegraphics[width=0.9\textwidth]{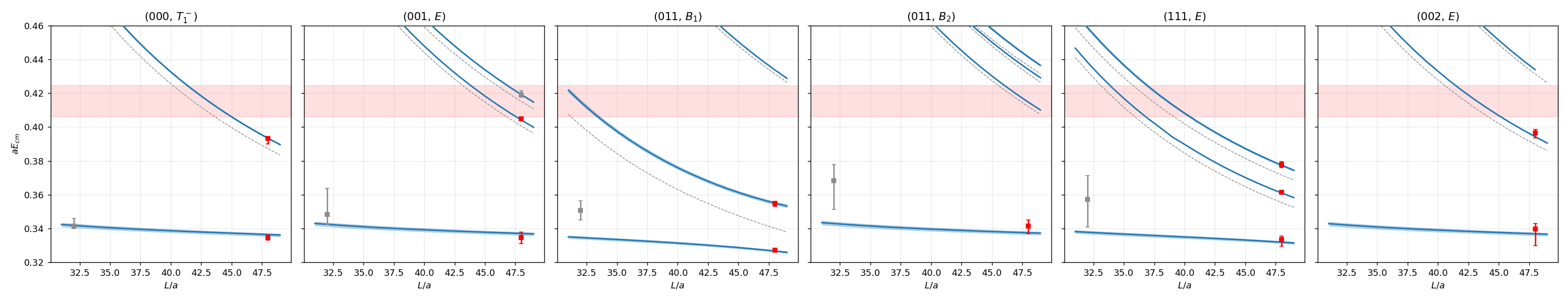}
	\caption{As in Fig.~\ref{fig:spectra1}, but for the F48P21 and
		F32P21 ensembles.}
	\label{fig:spectra2}
\end{figure}
\begin{figure}[H]
	\centering \includegraphics[width=0.8\textwidth]{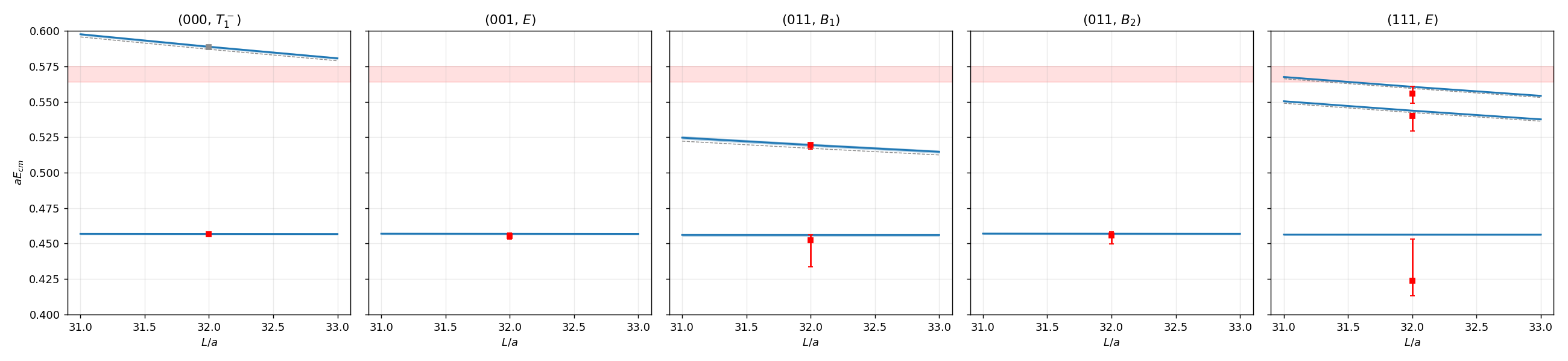}
	\caption{As in Fig.~\ref{fig:spectra1}, but for the C32P29
		ensemble.}
	\label{fig:spec-C32P29}
\end{figure}
\begin{figure}[H]
	\centering
	\includegraphics[width=0.8\textwidth]{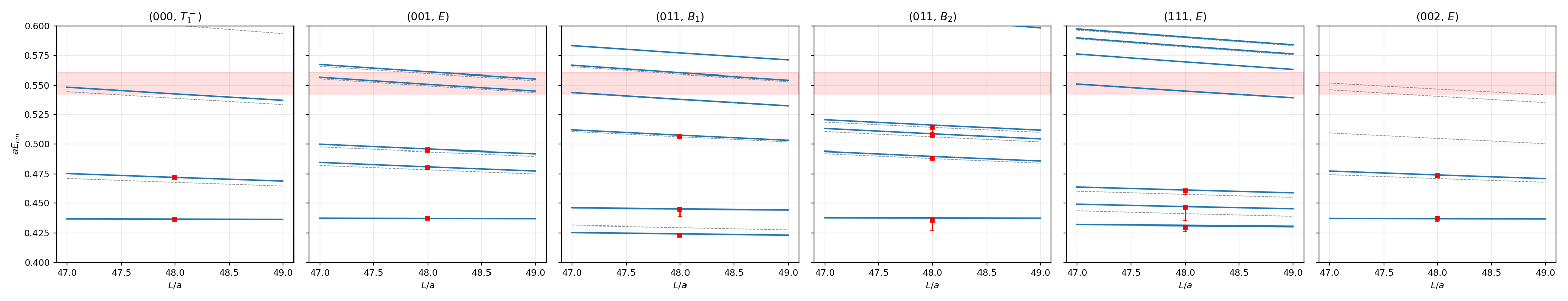}
	\caption{As in Fig.~\ref{fig:spectra1}, but for the C48P23
		ensemble.}
	\label{fig:spec-C48P23}
\end{figure}
\begin{figure}[H]
	\centering
	\includegraphics[width=0.8\textwidth]{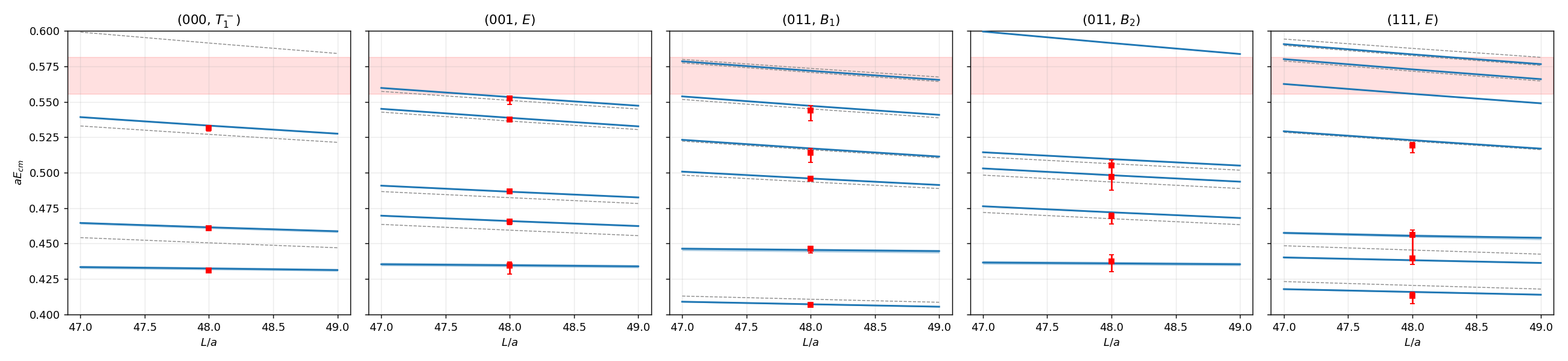}
	\caption{As in Fig.~\ref{fig:spectra1}, but for the C48P14
		ensemble.}
	\label{fig:spec-C48P14}
\end{figure}
\begin{figure}[H]
	\centering \includegraphics[width=0.8\textwidth]{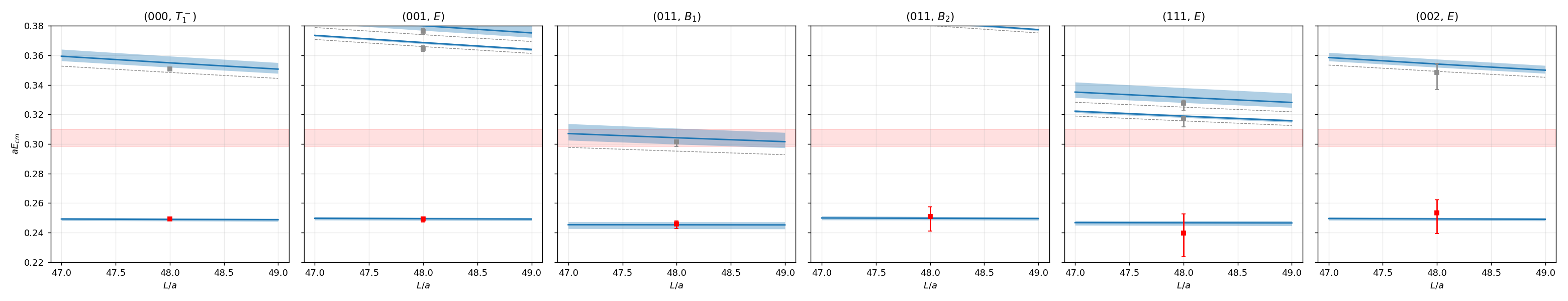}
	\caption{As in Fig.~\ref{fig:spectra1}, but for the H48P32
		ensemble.}
	\label{fig:spec-H48P32}
\end{figure}
} 

\section{Phase shifts determination}
\label{sec:ScatAna}
\subsection{L\"uscher's formula}\label{Lsa}
The L\"ushcer formula relates the finite-volume spectrum to the phase
shifts in the infinite space, i.e., the finite-volume spectrum are the
solution of the L\"uscher quantization condition (QC)
\begin{equation}
	\operatorname{ d e t } \left(  { 1 } + i \mathcal T \left( { 1 } + i \mathcal { M } ^ { \boldsymbol { P } } \right) \right) = 0 .
\end{equation}
for a two-particle system with total momentum $\boldsymbol P$. The
$\mathcal T$ matrix is diagonal in the $\ell m$ indices, i.e.,
\begin{equation}
	\mathcal T_{\ell m,\ell^\prime m^\prime} = T^{\ell} \delta_{\ell,\ell^\prime}\delta_{m,m^\prime}~.
\end{equation}
The elements of $\mathcal M^{\boldsymbol P}$ for $l,l^\prime \leq 1$
are given in Ref.~\cite{Leskovec:2012gb}
\begin{widetext}
	\begin{align}
		\label{M_gen}
		                      & \left({\cal M}^{\vec{P}}_{\ell m, \ell' m'}\right) = \bordermatrix{~ & 0 \phantom{\mbox{-}}0        & 1 \phantom{\mbox{-}}0   & 1 \phantom{\mbox{-}}1             & 1\mbox{-}1\cr
		0 \phantom{\mbox{-}}0 & \omega_{00}                                                          & i\sqrt{3}\omega_{10}         & i \sqrt{3}\omega_{11}   & i\sqrt{3}\omega_{1 \mbox{-}1} \cr
		1 \phantom{\mbox{-}}0 & -i\sqrt{3}\omega_{10}                                                & \omega_{00}+2\omega_{20}     & \sqrt{3}\omega_{21}     & \sqrt{3}\omega_{2\mbox{-}1} \cr
		1 \phantom{\mbox{-}}1 & i \sqrt{3}\omega_{1\mbox{-}1}                                        & -\sqrt{3}\omega_{2\mbox{-}1} & \omega_{00}-\omega_{20} & -\sqrt{6} \omega_{2\mbox{-}2} \cr
		1 \mbox{-}1           & i \sqrt{3} \omega_{11}                                               & -\sqrt{3}\omega_{21}         & -\sqrt{6}\omega_{22}    & \omega_{00}-\omega_{20} \cr}\ ,
	\end{align}
\end{widetext}
with function $\omega_{\ell m}$
\begin{equation}
	\omega_ { \ell m } = \omega_ { \ell m } ^ { \boldsymbol{ P } } ( k , \, L ) = \frac { Z _ { \ell m } ^ { \boldsymbol { P } } \left( 1 ; ( k \frac { L } { 2 \pi } ) ^ { 2 } \right) } { \gamma \pi ^ { 3 / 2 } \sqrt { 2 \ell + 1 } ( k \frac { L } { 2 \pi } ) ^ { \ell + 1 } } \, .
\end{equation}

Here, $Z^{\boldsymbol P}_{\ell m}$ is the generalized zeta function
and $\gamma = E^{\boldsymbol P}/\sqrt{s}$ is the Lorentz boost factor.
Incorporating the symmetries of the little groups $LG(\boldsymbol P)$,
the matrix $\mathcal M^{\boldsymbol P}$ can be further simplified and
block-diagonalized, leading to different quantization conditions for
different little groups and irreps. In Table~~\ref{tab:qc}, we list
the quantization conditions for all $LG(\boldsymbol P)$ and irreps
that are used to determine the P-wave phase shift $\delta_1$.
\begin{table}[H]
	\centering \renewcommand{\arraystretch}{1.3}
	\begin{tabular}{|c|c|c|c|}
		\hline
		$\boldsymbol P$              & LG($\boldsymbol P)$        & irrep     & QCs                                                                                        \\
		\hline
		{$(0, 0, 0)$}                & {$O_h$}                    & $T_1^{-}$ & $\cot \delta_1=\omega_{00}$                                                                \\
		\hline {$(0, 0, n)$}         & {$C_{4 v}$}                & $E$       & $\cot \delta_1=\omega_{00}-\omega_{20}$                                                    \\
		\hline
		\multirow{2}{*}{$(n, n, 0)$} & \multirow{2}{*}{$C_{2 v}$} & $B_2$     & $\cot \delta_1=\omega_{00}+2 \omega_{20}$                                                  \\
		\cline{3-4}                  &                            & $B_1$     & $\cot \delta_1=\omega_{00}-\omega_{20}-\sqrt{6} \operatorname{Im}\left[\omega_{22}\right]$ \\
		\hline
		{$(n, n, n)$}                & {$C_{3 v}$}                & $E$       & $\cot \delta_1=\omega_{00}+i \sqrt{6} \omega_{22}$                                         \\
		\hline
	\end{tabular}
	\caption{L\"uscher quantization conditions for little groups
		and irreps used in this work.}
	\label{tab:qc}
\end{table}
In practice, we solve the QCs to obtain a set of finite-volume energy
levels $\{\tilde E_n^{\Lambda,\boldsymbol P,L}(\boldsymbol \alpha)\}$,
where $\boldsymbol \alpha$ denotes the parameters in the
parameterization of the phase shifts $\delta_1$. Those parameters
$\boldsymbol \alpha$ are determined by minimizing
\begin{equation}\label{chi}
	\chi^2 = \sum_L \left( \sum_{\Lambda, n,\boldsymbol P} \sum_{\Lambda^\prime, n^\prime,\boldsymbol P^\prime}\left(E^{\Lambda, \boldsymbol P ,L}_n - \tilde E^{\Lambda, \boldsymbol P,L}_n  \right) \left[ C^{-1}(L) \right]_{\Lambda,n,\boldsymbol P; \Lambda^\prime,n^\prime,\boldsymbol P^\prime} \left(E^{\Lambda^\prime, \boldsymbol P^\prime ,L}_{n^\prime} - \tilde E^{\Lambda^\prime, \boldsymbol P^\prime,L}_{n^\prime} \right) \right)~.
\end{equation}
Here, $\left[C^{-1}\right]$ is the inverse of the covariance matrix of
the energy levels $\{E^{\Lambda, \boldsymbol P ,L}_n\}$ determined on
the lattice. The fitted energy levels are chosen below the $K\eta$
threshold, to avoid the effects from the inelastic channels, and they
are shown in Fig.~~\ref{fig:spectra1}-\ref{fig:spec-H48P32} by red
points for all ensembles.
%
%
To proceed, the phase shift $\delta_1$ must be parameterized by
suitable energy-dependent function, then the amplitudes can be
analytically continued into the complex plane to search for
resonances.

\subsection{Parameterizations of the scattering amplitudes}\label{PSC}

According to unitarity condition, the partial-wave amplitude (PWA)
$T_\ell$ for $2\to2$ elastic scattering obeys the constraint
\begin{equation}\label{ImT}
	\operatorname{Im}T_\ell^{-1}(s) = -i\rho(s),\quad s>s_R ~ ,
\end{equation}
with two-body phase-space factor $\rho(s)$ and threshold $s_R$
\begin{equation}
	\rho(s)=\frac{\sqrt{(s-s_L)(s-s_R)}}{s},\quad s_L=(m_1-m_2)^2,\quad s_R=(m_1+m_2)^2.
\end{equation}
The $m_i, (i=1,2)$ denote the masses for scattering particles. The
$S_\ell$ matrix is defined as
\begin{equation}
	S_\ell(s)=1+2i\rho(s)T_\ell(s),\quad S_\ell S_\ell^\dagger =1,
\end{equation}
and the phase shift $\delta_\ell(s)$ is related to $S$ matrix as
\begin{equation}
	S_\ell(s)=e^{2i\delta_\ell(s)}~.
\end{equation}
Several approaches to construct a unitary scattering amplitude are
used subsequently for comparison.
\subsubsection{The $K$ matrix}
The $K-$matrix formalism is a simple and convenient method to
construct a unitarity PWA, which expresses the amplitude in terms of a
$K_\ell(s)$ function
\begin{equation}
	T_\ell^{-1}(s) = \frac{1}{(2k)^{2\ell}} K_\ell^{-1}(s)-i\rho(s)~.
\end{equation}
The center-of-mass momentum $k=\frac{\rho(s)\sqrt{s}}{2}$ is
introduced to ensure the behavior of PWA at threshold with angular
momentum $\ell$. The unitarity condition Eq.~\eqref{ImT} implies that
the $K_{\ell}(s)$ function is analytic above the threshold, i.e.,
$\operatorname{Im} K_\ell(s)=0$, for $s>s_R$. In the lattice
literature, the $K_\ell(s)$ function is usually expressed as a pole
term, a polynomial, or a combination of both. However, it is crucial
to realize that these parameterizations for $K_\ell(s)$ all neglect
the contributions from the left-hand cuts. Consequently, these
parameterizations may lead to an incorrect analytic structure on the
complex energy plane, although they may be capable of describing the
phase shifts well on the real axis.
In the $K$-matrix formalism, the relation between function $K_\ell(s)$
and phase shift $\delta_\ell(s)$ can be written as
\begin{equation}
	k^{2\ell+1}\cot\delta_\ell(s)=\frac{\sqrt{s}}{2^{2\ell+1}} K_\ell^{-1}(s)~.
\end{equation}
with
\begin{equation}\label{polep}
	K_\ell(s) = \frac{g_\ell^2}{m_\ell^2-s} + \sum_{n} c_{\ell,n} s^{n}~.
\end{equation}
The effective range expansion (ERE) is also widely used to
parameterize the $K_\ell$ function, which expresses the phase shifts
$\delta_l$ as
\begin{equation}
	k^{2\ell+1}\cot\delta_\ell(s) = \frac{1}{a_\ell} + \frac{r_\ell}{2} k^2+ \cdots,
\end{equation}
or the $K_\ell(s)$ function as
\begin{equation}
	K_\ell^{-1}(s)= \frac{2^{2\ell+1}}{\sqrt{s}}\left(\frac{1}{a_\ell} + \frac{r_\ell}{2} k^2 + \cdots \right)~,
\end{equation}
where the dots denote higher-order terms in $k^2$. In the following
analysis, we truncate the expansion at order $k^2$.

\subsubsection{Product representation (PKU representation) }
For the single channel elastic scattering, one can express the
$S_\ell$ matrix in terms of a product form~\cite{Zheng:2003rw}
\begin{equation}
	S_\ell(s) =  \Pi_i S_\ell^{p_i}(s) \cdot S_\ell^{cut}(s)~,
\end{equation}
where the product of $S_\ell^{p_i}$ factors represents the unitary
factor contributed by the isolated singularities of $S_\ell$ matrix,
which come from three possible state types: virtual states, bound
states and resonance states. For a virtual state at $s_v$, the factor
denoted by $S^v$ is given by
\begin{equation}
	S_\ell^v(s) = \frac{1 + i \rho(s) \frac{s}{s - s_L} \sqrt{\frac{s_v - s_L}{s_R - s_v}}}{1 - i \rho(s) \frac{s}{s - s_L} \sqrt{\frac{s_v - s_L}{s_R - s_v}}},\quad s_L < s_v <s_R~.
\end{equation}
The contribution from a bound state at $s_b$ denoted by $S^b$ is the
inverse of $S^v$ with $s_v$ replaced with $s_b$. A pair of resonances
at $z_r$ (having the positive imaginary part) and $z_r^*$, the $S^r$
has the form
\begin{equation}
	S_\ell^r ( s ) = \frac { M ^ { 2 } ( z _ { r } ) - s + i \rho ( s ) s G } { M ^ { 2 } ( z _ { r } ) - s - i \rho ( s ) s G } ~,
\end{equation}
with
\begin{equation}
	G = \frac{\text{Im}[z_r \rho(z_r)]}{\text{Re}[z_r \rho(z_r)]},\quad
	M^2(z_r) = \text{Re}[z_r] + \frac{\text{Im}[z_r] \text{Im}[z_r \rho(z_r)]}{\text{Re}[z_r \rho(z_r)]}.
\end{equation}
The remaining part is the left-hand cut contribution denoted by
$S^{cut}$, which can be parameterized as
\begin{equation}
	S_\ell ^ { c u t } = \exp [ 2 i \rho ( s ) f ( s ) ] \ ,
\end{equation}
where the $f(s)$ satisfies the following dispersion relation
neglecting the inelastic effects
\begin{equation}
	f(s) = f(s_0) + \frac{(s - s_0)}{2\pi i} \int_L \frac{\text{disc}_L f(z)}{(z - s)(z - s_0)} dz
	~,
\end{equation}
with $L$ denoting the integration path on the left-hand cut. In
general, the cut contributions can be evaluated approximately from a
low-energy effective field theory, such as $\chi$PT. The product
representation separates the partial waves into various unitary
factors contributed either from poles or branch cuts, such that the
corresponding phase shifts are additive and each phase shift
contribution has a definite sign, which makes possible the
disentanglement of hidden poles from a background. In addition, the
consistency of the product representation with crossing symmetry is
also examined in Refs.~\cite{Guo:2007ff,Guo:2007hm}.

This method has been successfully applied to investigate the
$\pi\pi$~\cite{Zhou:2004ms}, $\pi K$~\cite{Zheng:2003rw} and
$\pi N$~\cite{Wang:2018nwi} scatterings at physical pion mass. In
addition to determining the positions of the $\sigma$ and $\kappa$,
the method also reveals the existence of a hidden pole below the
threshold of $\pi N$, which has also been corroborated in the
Roy-Steiner equation analysis~\cite{Cao:2022zhn, Hoferichter:2023mgy}
and other unitarization methods~\cite{LiQuZhi:2021nrq}.

The key point in the application of the product representation is to
estimate the contributions from the left-hand cuts. Based on the fact
that function $f(s)$ only contains the left-hand cuts, which means it
is analytic on the whole complex plane except on the left-hand cuts,
we expand the $f(s)$ function in terms of a suitably constructed
conformal mapping variable $\omega(s)$
\begin{equation}\label{fwn}
	f(s) = \sum_n C_n  \omega^n(s)~.
\end{equation}
with
\begin{equation}
	\omega ( s ) = - \frac { \left( \, \sqrt { s } -  \sqrt { s _ { E } } \right) \left(\sqrt { s } \, \sqrt { s _ { E } } + s _ { - } \right) } { \left( \, \sqrt { s } + \, \sqrt { s _ { E } } \right) \left( \, \sqrt { s } \, \sqrt { s _ { E } } - s _ { - } \right) } \: , \quad s_-=m_2^2-m_1^2.
\end{equation}
The conformal transformation maps the left-hand cut onto the unit
circle~\cite{Frazer:1961zz}. Since the series in Eq.~\eqref{fwn}
converges within the unit disk in the $\omega$ plane, it converges on
the entire $s$ plane, except on the left-hand cuts. Furthermore, in
the analytic region such as the region on the right-hand cut of the
$S$-matrix, it converges more rapidly than the ordinary power series
in terms of $s$.

Finally, the partial-wave $S$ matrix reads
\begin{equation}
	S_\ell(s) = \Pi_i S_\ell^{p_i}(s) \cdot e^{2i\rho(s)\sum_{n} C_n \omega^n(s)}~.
\end{equation}
For the $P-$wave, since the phase shifts $\delta_1(k)$ behaves like
\begin{equation}
	\delta_1(k) \sim  O(k^3)~,
\end{equation}
near the threshold, when expanded as a power series of $k$, the $k^1$
term should vanish\footnote{The phase shifts is the odd function in
	terms of $k$. }. Expanding the parameterization \eqref{fwn} in terms
of $k$, the contributions of resonance, virtual state , bound state
and the left-hand cuts to the coefficients of the $k^1$ read
\begin{equation}
	A_{\mathrm{res.}}(z_{r}) = \frac{2(m_\pi + m_K)G_{r}}{M_{r}^{2} - s_{R}},
\end{equation}
\begin{equation}
	A_{v.s.}(s_v) = \sqrt{\frac{s_v - s_L}{s_R - s_v}} \frac{m_\pi + m_K}{2m_\pi m_K}~,
\end{equation}
\begin{equation}
	A_{b,s}(s_b) = -\sqrt{\frac{s_b - s_L}{s_R - s_b}} \frac{m_\pi + m_K}{2m_\pi m_K}~,
\end{equation}
\begin{equation}
	A _ { b g } = \frac { 2 f ( s _ { R } ) } { m_\pi+m_K} ~,
\end{equation}
respectively. The constraint of vanishing $k^1$ term implies
\begin{equation}\label{CC}
	\sum_{b} A_{b.s.}(s_{b}) + \sum_{v} A_{v.s.}(s_{v}) + \sum_{r} A_{res.}(z_{r}) + A_{bg} = 0~.
\end{equation}
In the minimization~Eq.\eqref{chi}, we analytically continue the
$k^{2l+1}\cot\delta_\ell$ below the threshold to fit the energy levels via
\begin{equation}
	k^{2\ell+1}\cot\delta_\ell = \frac{ k^{2\ell}\sqrt{s}}{2}\left( \frac{1}{T_\ell(s)} +  i\rho(s)  \right)~.
\end{equation}

\subsection{Numerical results}\label{numre}
Taking advantage of the fact that the QCs only contain the $P$-wave in
the irreps: $T_1^-,E,B_1$ and $B_2$ when ignoring the higher partial
waves ( $\ell\geq 2$), we analyze the $P$-wave phase shift and the
associated resonance structure.
Three distinct model functions are used to describe $\delta_1(s)$: the
$K$ matrix in Eq.~\eqref{polep} (labeled by ``BW") , ERE, and the
product representation (labeled by ``PR") involving one conformal
variables and one resonance state. The later contains only two free
parameters, taking into account the constraint Eq.~\eqref{CC}.

\begin{table}[H]
	\centering \renewcommand{\arraystretch}{1.8}
	\resizebox{\textwidth}{!}{
		\begin{tabular}{|c|c|c|c|c|}
			\hline
			ensemble & model & parameters (unit of the lattice spacing)                             & pole position (MeV)                              & $\chi^2/\mathrm{dof}$ \\
			\hline
			\multirow{3}{*}{F48P21}
			         & BW    & $g^2=0.188^{+0.028}_{-0.029}$, $m^2=0.1130^{+0.0010}_{-0.0013}$      & $854.7^{+3.7}_{-4.7}-i\,12.8^{+2.0}_{-2.2}$      & 0.56                  \\
			         & ERE   & $1/a_1=0.0085^{+0.0015}_{-0.0011}$, $r_1/2=-1.015^{+0.129}_{-0.184}$ & $854.2^{+3.7}_{-4.7}-i\,12.1^{+1.9}_{-2.0}$      & 0.55                  \\
			         & PR    & $z=0.1126^{+0.0010}_{-0.0012}+i\,0.0034^{+0.0005}_{-0.0006}$         & $854.9^{+3.8}_{-4.7}-i\,13.0^{+2.0}_{-2.2}$      & 0.57                  \\
			\hline
			 F48P30        & BW    & $g^2=0.132^{+0.040}_{-0.044}$, $m^2=0.1235(5)$                       & $895.2^{+1.8}_{-2.0}-i\,2.8^{+0.7}_{-0.8}$       & 0.70                  \\
			 \&        & ERE   & $1/a_1=0.0063^{+0.0030}_{-0.0015}$, $r_1/2=-1.512^{+0.338}_{-0.678}$ & $895.7^{+1.7}_{-1.7}-i\,2.7^{+0.7}_{-0.8}$       & 0.73                  \\
			 F32P30        & PR    & $z=0.1235(5)+i\,0.0008^{+0.0002}_{-0.0003}$                          & $895.3^{+1.8}_{-1.8}-i\,2.9^{+0.7}_{-0.9}$       & 0.70                  \\
			\hline
			\multirow{3}{*}{C48P23}
			         & BW    & $g^2=0.134^{+0.017}_{-0.016}$, $m^2=0.1912^{+0.0007}_{-0.0008}$      & $819.5^{+1.4}_{-1.7}-i\,5.8^{+0.8}_{-0.8}$       & 0.91                  \\
			         & ERE   & $1/a_1=0.0196^{+0.0027}_{-0.0028}$, $r_1/2=-1.836^{+0.262}_{-0.267}$ & $819.2^{+1.5}_{-1.7}-i\,5.5^{+1.0}_{-0.8}$       & 0.95                  \\
			         & PR    & $z=0.1910^{+0.0006}_{-0.0008}+i\,0.0027(4)$                          & $819.5^{+1.4}_{-1.6}-i\,5.9^{+0.8}_{-0.8}$       & 0.90                  \\
			\hline
			\multirow{3}{*}{C48P14}
			         & BW    & $g^2=0.181^{+0.021}_{-0.025}$, $m^2=0.1931^{+0.0018}_{-0.0022}$      & $822.4^{+3.8}_{-4.6}-i\,12.7^{+1.4}_{-1.8}$      & 0.81                  \\
			         & ERE   & $1/a_1=0.0214^{+0.0031}_{-0.0021}$, $r_1/2=-1.434^{+0.126}_{-0.209}$ & $821.9^{+3.6}_{-4.7}-i\,12.2^{+1.3}_{-1.7}$      & 0.82                  \\
			         & PR    & $z=0.1924^{+0.0017}_{-0.0021}+i\,0.0060^{+0.0007}_{-0.0009}$         & $822.5^{+3.6}_{-4.6}-i\,12.9^{+1.5}_{-1.9}$      & 0.71                  \\
			\hline
			\multirow{3}{*}{C32P29}
			         & BW    & $g^2=0.034^{+0.046}_{-0.032}$, $m^2=0.2077^{+0.0007}_{-0.0008}$      & $854.51^{+1.54}_{-1.67}-i\,0.52^{+0.64}_{-0.50}$ & 0.94                  \\
			         & ERE   & $1/a_1=0.0269^{+0.0011}_{-0.0072}$, $r_1/2=-4.999^{+1.360}_{-0.001}$ & $854.13^{+1.26}_{-1.36}-i\,0.73^{+0.26}_{-0.04}$ & 1.05                  \\
			         & PR    & $z=0.2077(8)+i\,0.0003(3)$                                           & $854.46^{+1.55}_{-1.69}-i\,0.54^{+0.65}_{-0.51}$ & 0.92                  \\
			\hline
			\multirow{3}{*}{H48P32}
			         & BW    & $g^2=0.242^{+0.227}_{-0.189}$, $m^2=0.0602^{+0.0011}_{-0.0016}$      & $931.4^{+8.3}_{-13.1}-i\,6.4^{+2.4}_{-3.3}$      & 0.68                  \\
			         & ERE   & $1/a_1=0.0011^{+0.0025}_{-0.0006}$, $r_1/2=-0.510^{+0.239}_{-0.986}$ & $931.9^{+8.7}_{-13.0}-i\,6.1^{+2.4}_{-3.7}$      & 0.66                  \\
			         & PR    & $z=0.0601^{+0.0012}_{-0.0017}+i\,0.0008^{+0.0003}_{-0.0006}$         & $932.6^{+9.0}_{-13.1}-i\,6.0^{+2.7}_{-4.5}$      & 0.68                  \\
			\hline
		\end{tabular}}
	\caption{The fit results obtained using the three parametrizations for
		all ensembles used in this work. The pole positions and the values
		of $\chi^2/\text{dof}$ are also presented. The central values and
		uncertainties of the parameters correspond to the median values and the 16th and 84th percentiles, respectively. Uncertainties in  parentheses  indicate symmetric errors.   The last
		column lists the $\chi^2/\text{dof}$ values for the median values
		across samples.}	\label{tab:fitrP}
\end{table}

\begin{figure}[H]
	\centering
	\includegraphics[width=\linewidth]{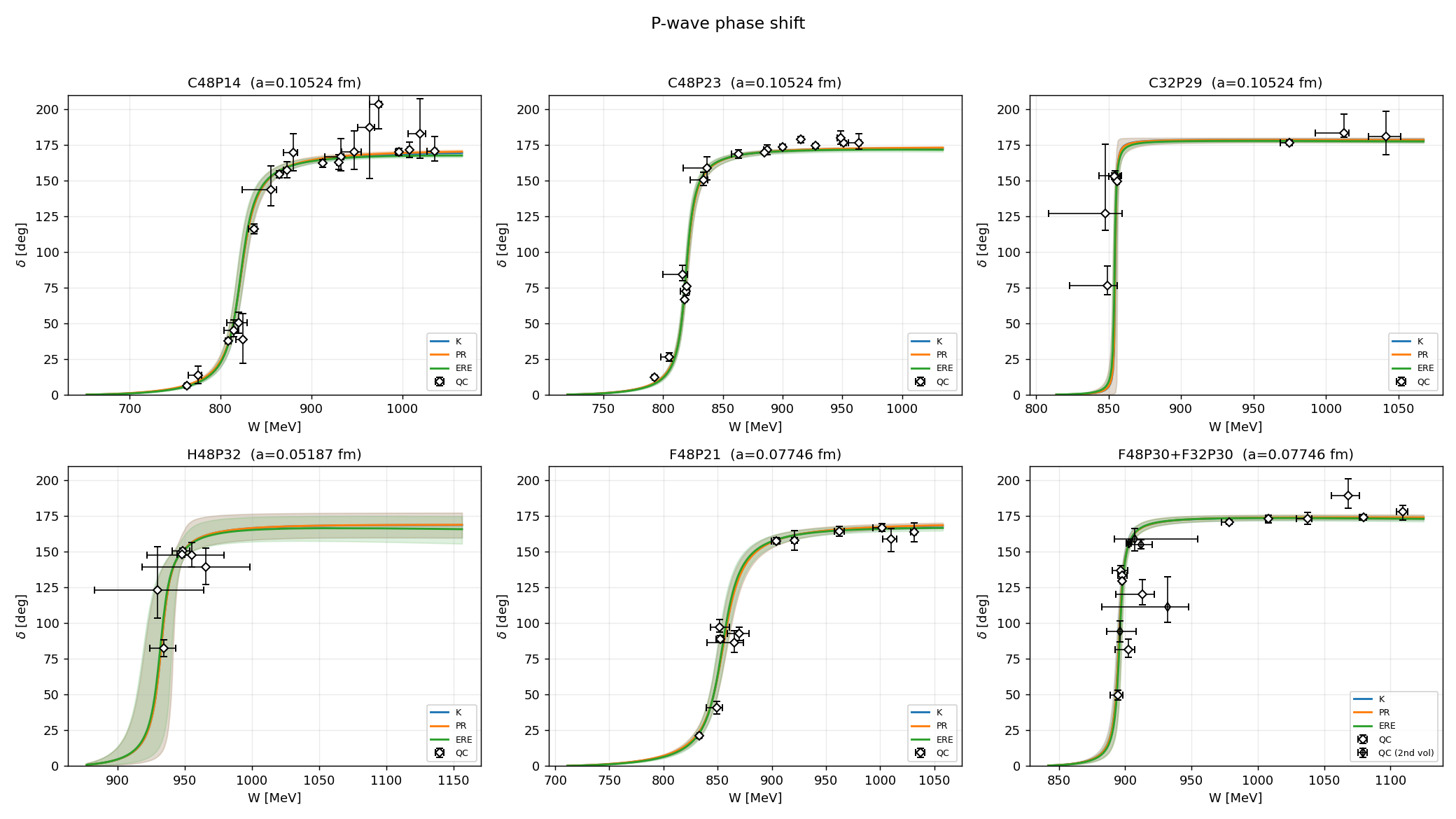}
	\caption{The P-wave $K\pi$ phase shift $\delta_1$ as a
		function of the center-of-mass energy($W$). The results from
		three different parametrizations are shown as blue(PR),
		orange(ERE) and green(BW) dashed lines, respectively. }

	\label{fig:phase-shift}
\end{figure}

The fit results, listed in Tab.~\ref{tab:fitrP}, show that all models
yield consistent spectrum within uncertainties.
The energy dependence of the phase shift from the three parameterizations are shown in
Fig.~\ref{fig:phase-shift}, where all phase-shift curves exhibit a
rapid increase, implying that there is a narrow resonance.
The resonance corresponds to the pole singularity of PWA on the
unphysical Riemann sheet. For a narrow resonance (the decay width much
smaller than the mass), the pole position $s_0$ can be related to the
mass $m_R$ and decay width $\Gamma$ via $\sqrt{s_0}=m_R+ i \Gamma/2$.

For all ensembles, all the three distinct parameterizations are able
to describe the finite-volume spectrum feature with a single pole
close to the real axis on the second Riemann sheet of complex energy
plane. The pole positions are also summarized in
Tab.~\ref{tab:fitrP}. Notably, the fitting parameters for the ``PR"
model correspond directly to the resonance position.

\begin{figure}[H]
	\centering
	\includegraphics[width=\linewidth]{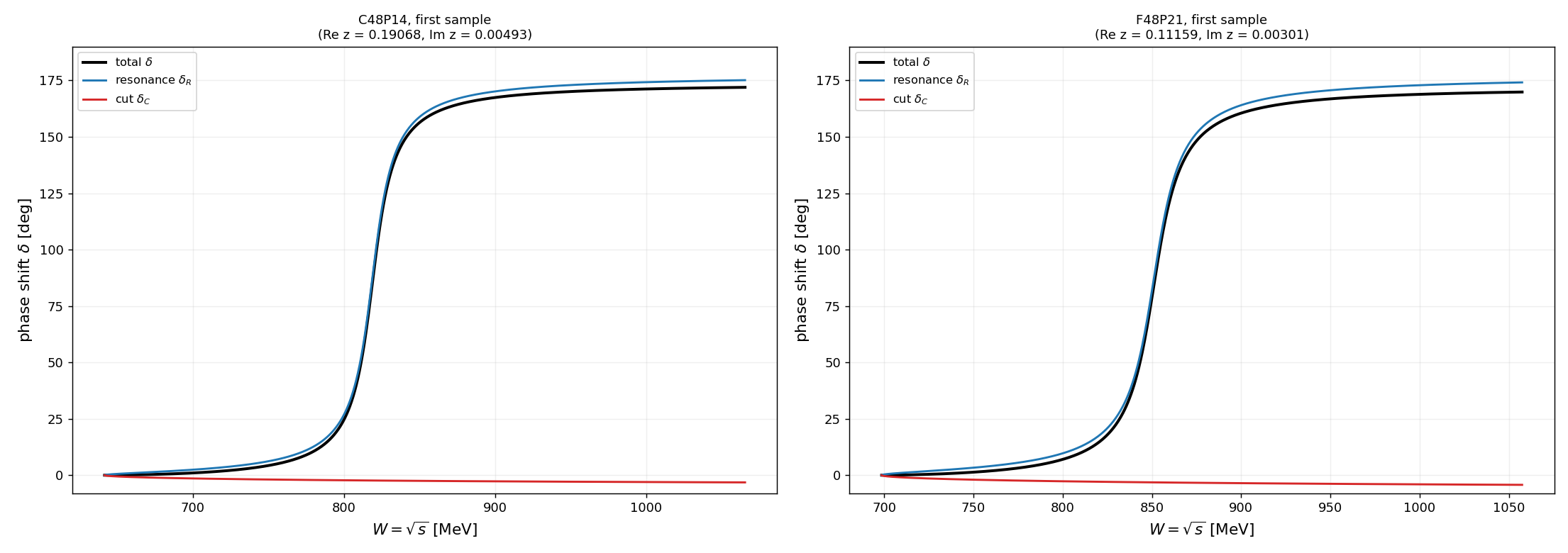}
	\caption{The PKU decomposition for the phase shifts of C48P14
		(left) and F48P21 (right) ensembles. The red dotted line
		represents the left-hand cut contribution, the black line
		shows the resonance contribution, and the blue dashed line
		denotes the total phase shift. { As an example,  we only  use the parameter values $z$ from the first bootstrap sample to calculate the total phase shifts and the contributions from resonance and cuts, respectively. }}
	\label{fig:PKU-decom}
\end{figure}

As mentioned previously, in the ``PR" model, the total phase shifts
can be decomposed into different contributions from the poles and
left-hand cuts. As an example, the total phase shifts for ensembles
F48P21 and C48P14 are decomposed into the resonance $K^*(892)$ and the
left-hand cut contributions in Fig.~\ref{fig:PKU-decom}. The later is
expressed as
\begin{equation}
	\delta^{cut}(s) = \rho(s) f(s)=\rho(s) C_0~.
\end{equation}
where the $C_0$ is obtained from Eq.~\eqref{CC}.

As shown in Fig.~\ref{fig:PKU-decom}, the cut contributions are
negative and small, while the resonance contributes positive and large
phase shifts, and their summation gives the total phase shifts. In
both scenarios, the cut contributions are negative, which is
consistent with the analysis in the $S$- and $P$-waves of the
$\pi\pi$, $\pi K$, and $\pi N$ systems with physical pion
mass~\cite{Xiao:2000kx,Zhou:2004ms,Wang:2018nwi}. However, compared
with the negligible cut contributions in the $P$-wave (see
Fig.~\ref{fig:PKU-decom} and Fig.3 in Ref.~\cite{Zhou:2004ms} for the
$IJ=11$ channel of $\pi\pi$ scattering), the negative cut
contributions in the $S$-wave are typically too negative to allow for
positive total phase shifts without resonance contributions. To cancel
the cut contributions and achieve positive total phase shifts,
resonances such as the $\sigma$ (for the $I=J=0$ channel of $\pi\pi$
scattering~\cite{Xiao:2000kx}), $\kappa$ (for the $I=1/2, J=0$ channel
of $\pi K$ scattering~\cite{Zhou:2004ms}), and $N^*(920)$ (for the
$I=1/2, J=1/2$ channel of $\pi N$ scattering~\cite{Wang:2018nwi}) must
exist. We would leave the similar analysis for the $S$-wave $\pi K$
system from the lattice data with different pion mass for a future
work.

\begin{figure}[htpb]
	\centering \includegraphics[width=0.75\linewidth]{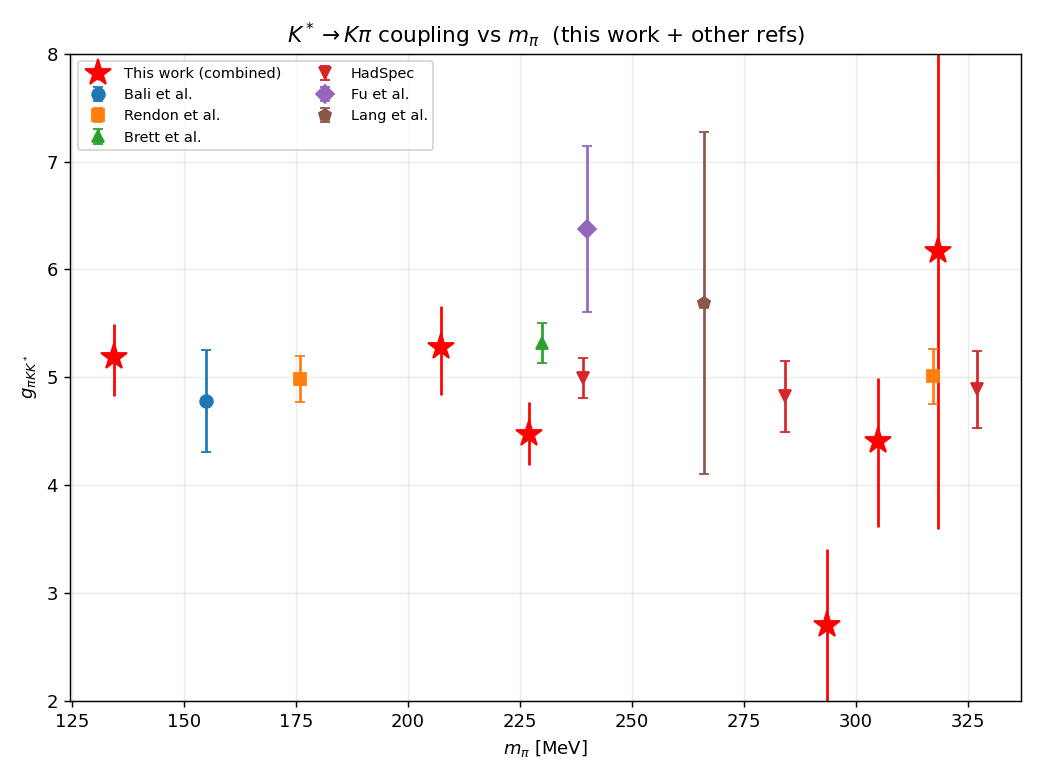}
	\caption{The coupling   $g_{\pi KK^*}$ for the different pion masses in this work compared with other lattice QCD results~\cite{Wilson:2019wfr,Fu_2012,Bali:2015gji,Brett:2018jqw,Lang_2012,Rendon:2020rtw}.
	}
	\label{fig:coupling}
\end{figure}

Another quantity of phenomenological interest is the $\pi KK^* $
coupling constant $g_{\pi K K ^*}$, which is quite well approximated
by a constant as the quark mass varies\cite{Nebreda:2010wv}. The
coupling constant can be extracted from the decay width using the
equation
\begin{equation}
	\Gamma = \frac{g_{\pi K K^*}^2}{6\pi}  \frac{k_{*}^3}{m_R^2}~,
\end{equation}
where $k_*$ is the scattering momentum with $s=m_R^2$. Our results of
$g_{\pi K K^*}$ from the AIC weighted mean of the three
parametrizations are compared with other lattice results in
Fig.~\ref{fig:coupling}.

\subsection{Extrapolations}\label{extr}
Six solutions have been obtained for $K^*(892)$ resonance at different
pion masses and finite lattice spacings, which can be used for
extrapolations to physical quark masses and continuum limit.
Based on the formulas
{ 
\begin{equation}\label{eq:extrop}
	\begin{split}
		\mathrm{Re}\left(\sqrt{s_0}\right) = b^r_0+b^r_1 ( m_{\pi,r} ^2 -1) + b^r_2 (m_{K,r}^2 - 1)+ b^r_3 a^2, \\
		\mathrm{Im}\left(\sqrt{s_0}\right) = b^i_0+b^i_1 (m_{\pi,r} ^2 - 1) + b^i_2 (m_{K,r}^2 -1)+ b^i_3 a^2,  \\
	\end{split}
\end{equation}
with dimensionless quantities
$m_{\pi/K,r} = m_{\pi/K}/m_{\pi/K}^{\text{phy}}$, we fit the
real and imaginary parts of the $K^*(892)$ poles across all
ensembles for each  parametrization models. The parameters $b^r_0$($b^i_0$) is the real(imaginary) part of the pole positions at physical pion and kaon masses and continuum limit. 
} 

\renewcommand{\arraystretch}{1.6}


\begin{table}[htbp]
	\centering
	\begin{tabular}{|c|c|c|c|c|c|c|}
		\hline
		model                & part & $b_0$ [MeV] & $b_1$ [MeV]            & $b_2$ [MeV]             & $b_3$ [MeV/fm$^2$]    & median $\chi^2/\mathrm{dof}$ \\
		\hline
		\multirow{2}{*}{BW}   & Re   & $882.05^{+7.42}_{-10.14}$ & $8.30^{+1.46}_{-1.32}$ & $153.8^{+34.9}_{-44.8}$ & $-6209^{+972}_{-709}$ & 1.38                         \\
		                     & Im   & $-22.72^{+3.40}_{-3.32}$ & $3.65^{+0.47}_{-0.56}$ & $-3.7^{+14.9}_{-13.4}$  & $962^{+279}_{-288}$   & 1.31                         \\
		\multirow{2}{*}{PR}   & Re   & $882.80^{+7.72}_{-10.21}$  & $8.32^{+1.44}_{-1.21}$ & $156.6^{+35.6}_{-45.8}$ & $-6297^{+985}_{-738}$ & 1.37                         \\
		                     & Im   & $-22.70^{+3.62}_{-3.50}$ & $3.69^{+0.50}_{-0.58}$ & $-2.4^{+15.7}_{-14.4}$  & $940^{+293}_{-318}$   & 1.06                         \\
		\multirow{2}{*}{ERE}  & Re   & $881.73^{+7.34}_{-10.25}$  & $8.46^{+1.47}_{-1.15}$ & $156.2^{+35.0}_{-46.2}$ & $-6253^{+997}_{-726}$ & 1.19                         \\
		                     & Im    & $-21.20^{+3.21}_{-3.10}$ & $3.43^{+0.41}_{-0.51}$ & $-4.1^{+13.9}_{-12.5}$  & $871^{+263}_{-275}$   & 1.26                         \\
		\hline
	\end{tabular}
	\caption{Fit results of the extrapolations to physical quark masses and continuum limit using Eq.~\ref{eq:extrop}. $b_0$ is the extrapolated pole positions at physical quark masses and continuum limit. }
	\label{tab:pole_params}
\end{table}

\renewcommand{\arraystretch}{1.0}{


The fit results are listed in Tables~\ref{tab:pole_params}
for each parametrization model. The coefficients of $m_\pi^2$ in the real
and imaginary parts of $K^*(892)$ indicate that the mass of the resonance
increases as $m_\pi$ increases, while its decay width decreases, as shown in
Fig.~\ref{fig:3d-re-im}. The latter behavior is rooted in the reduction of
the two-body phase space, because the coupling $g_{\pi K K^{*}}$ remains
constant, as shown above. 

{ 
To further isolate the lattice-spacing dependence, we define the  partial residuals of pole positions (with re the   $m_{\pi}$ and $m_{K}$ effects removed) as
\begin{equation}
	\label{eq:3}
	f^{\text{Im/Re}}_{\Delta} = b_0^{r/i}+ b^{r/i}_{3} a^{2} ~,
\end{equation}
and plot them in  Fig.~\ref{fig:residual-re-im}. These residuals clearly  show that discretization effects can be well described  by a linear function of $a_{r}^{2}$.

\begin{figure}[H]
	\centering
	\includegraphics[width=0.95\textwidth]{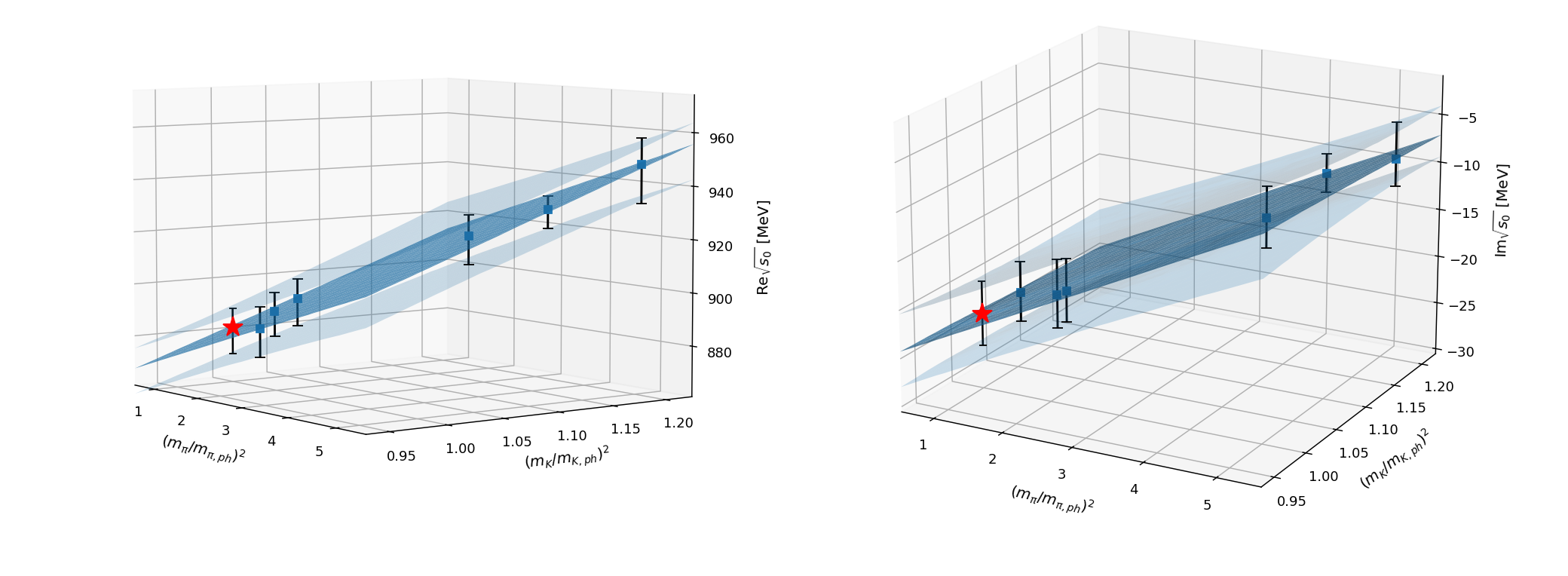}
	\caption{Real (left) and imaginary (right) parts of the pole positions as functions of $m_{\pi,r}^{2}$ and $m_{K,r}^{2}$, with the artificial lattice-spacing effects removed. The surfaces indicate the upper and lower bounds. Blue points correspond to the pole positions with discretization effects subtracted, while red points denote the extrapolated values. As an example, only the
		BW model is shown.}
	\label{fig:3d-re-im}
\end{figure}
\begin{figure}[H]
	\centering
	\includegraphics[width=0.95\textwidth]{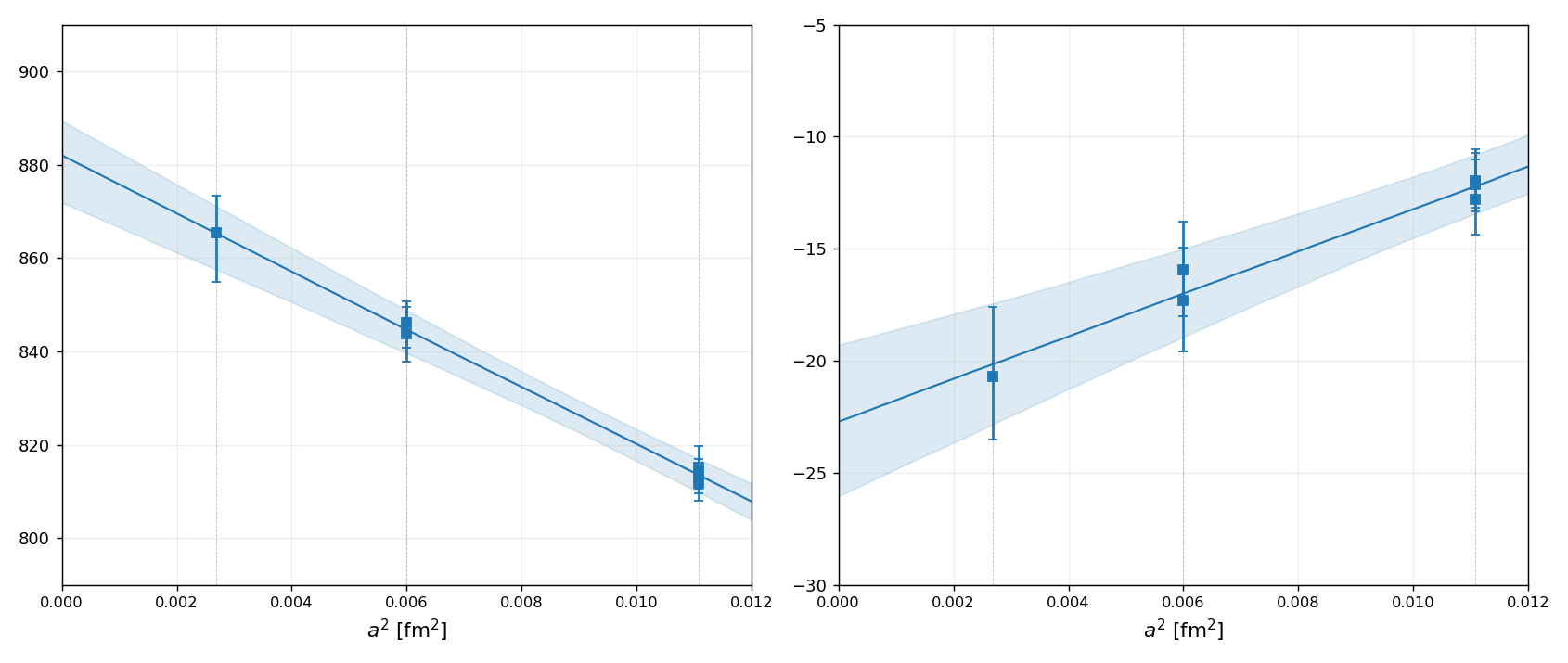}
	\caption{Partial residual plots for the pole position. The $y$-axis represents the partial residuals, isolating the component of the pole position not accounted for by   $m_{\pi,r}^{2}$ and $m_{K,r}^{2}$ (left: real part; right: imaginary part). The blue bands  indicate the theoretical trend based on the estimated slopes $b_{3}^{r/i}$. The points are given by Re(Im)$\sqrt{s_{0}} - b^{r(i)}_{0} - \sum_{j=1}^{2} b^{r(i)}_{j}(x_{j} - \bar x_{j}) $ for $x_{j}   = (m_{\pi,r}^{2}, m^{2}_{K,r})$. As an example, only the
		BW model is shown.}
	\label{fig:residual-re-im}
\end{figure}

Combining the parametrization of \(\mathrm{Re}(\sqrt{s_0})\) in Eq.~(\ref{eq:extrop}) with the leading-order GMOR relation~\cite{Okubo:1961jc}, we obtain the quark-mass dependence of the \(K^*(892)\) pole position:
\begin{align}
	\mathrm{Re}\bigl(\sqrt{s_0}\bigr)\big|_{a^2\rightarrow 0}
	 & = \sqrt{s_0}^{\mathrm{phy}}
	+\left(\frac{b_1^r}{m^{\mathrm{phy}}_{l}}+\frac{b_2^r}{m^{\mathrm{phy}}_{l}+m^{\mathrm{phy}}_{s}}\right)(m_l-m_l^{\mathrm{phy}})
	+\frac{b_2^r}{m^{\mathrm{phy}}_{l}+m^{\mathrm{phy}}_{s}}(m_s-m_s^{\mathrm{phy}}), \nonumber                               \\
	 & = \sqrt{s_0}^{\mathrm{phy}}+(m_l-m_l^{\mathrm{phy}})4.07^{+0.40}_{-0.43}+(m_s-m_s^{\mathrm{phy}})1.57^{+0.36}_{-0.46},
\end{align}
where \(m_l^{\mathrm{phy}}=3.37(4)\) MeV and \(m_s^{\mathrm{phy}}=93.4(1.0)\) MeV~\cite{CLQCD:2026jyw} in the \(\overline{\mathrm{MS}}\) scheme at 2 GeV. The resulting coefficients indicate a significantly stronger dependence on the light quark mass than on the strange quark mass. The light-quark coefficient \(4.07^{+0.40}_{-0.43}\) is comparable to the corresponding coefficient for the \(\rho\) meson evaluated at the same scale, \(3.9(6)(7)\)~\cite{Wang:2025hew}.

The extrapolated pole positions from the three parameterizations are combined using AIC weights to give our final results. Specifically, let $s_{b,m}$ denote the extrapolated pole positions obtained for bootstrap sample $b$ and parameterization $m$(m = "BW", "ERE", "PR"). For each $b$ and $m$,  an AIC weight $\omega_{b,m}$, similar to Eq.~\ref{eq:AIC},  is computed. Then the weighted mean $\hat{s}_{b} = \sum_{m} \omega_{b,m} s_{b,m} $ is computed for each bootstrap sample. The median and 16/84th percentile intervals of $\{\hat{s}_{b}\}_{b=0}^{9999}$ are then taken as the central value and uncertainty; this uncertainty combines the statistical uncertainty and systematic uncertainty arising from the choice of the fit windows, which is the first error quoted in our final results, Eq.~\ref{eq:2}. The systematic error due to different parameterization is estimated as $\sigma = \sum_m \omega_{0,m}( s_{0,m} - \hat{s}_{0} )^2$, which is the second error quoted in Eq.~\ref{eq:2}. 

Finally, We obtain  the pole position of $K^*(892)$
at the physical pion (139.57MeV) and kaon (493.677MeV)
masses at continuum limit:
\begin{equation}
	\label{eq:2}
	\sqrt{s_0}^{phy} = [882.10\binom{+7.36}{-10.29}(0.30) -i22.73(3.32)(0.63)] \mathrm{MeV}.
\end{equation}
} 

} 

Our extrapolation result are consistent with the values in the PDG and
previous lattice analysis at the physical pion mass, as shown in
Tab.~\ref{tab:polec}. 
{ 
We emphasize that the uncertainties quoted in our results have limitations. The systematic uncertainty associated with the choice of fit windows are not rigorously investigated due to the computational cost of fitting Lüscher's formula. Systematic uncertainties arising from the extrapolation ansatz, finite-volume correstions, isospin breaking, and other effects are not included in our error budget.
} 

\begin{table}[H]
\renewcommand{\arraystretch}{1.5} 
	\centering
	\begin{tabular}{|c|c|}
		\hline
		                                     & pole position (MeV)                                                                            \\
		\hline
		This work                            & $ 882.10\binom{+7.36}{-10.29}(0.30) -i22.73(3.32)(0.63)$ \\
		\hline
		PDG~\cite{ParticleDataGroup:2024cfk} & $ 890(14)-i26(6)   $                                                                           \\
		\hline
		Boyle et al.~\cite{Boyle:2024grr}    & $893(2)_{\text{stat}}(54)_{\text{sys}}-i 26(1)_{\text{stat}}(6)_\text{sys}$                    \\
		\hline
	\end{tabular}
	\caption{Comparison of the pole position from this work to the
		ones from PDG and previous lattice calculations at the
		physical pion mass. }
	\label{tab:polec}
\end{table}
\section{Conclusions and Outlooks}\label{COt}%
In this work, we study the P-wave $K\pi$ scattering using eight
$N_f=2+1$ Wilson-Clover ensembles with various pion masses and lattice
spacings. A large number of finite-volume energy levels are obtained
by using both rest frame and moving frames. The scattering phase
shifts are determined via L\"uscher's finite volume method with three
different parametrizations of the scattering amplitude.  The phase
shifts consistently exhibit a narrow resonance structure, and poles
corresponding to the $K^*(892)$ are found for all parameterizations
and all ensembles. The pole positions are then extrapolated to the
physical pion and kaon masses and to the continuum limit. The
extrapolated pole of the $K^*(892)$ is located at
$\sqrt{s_0}=[881.0(10.75)_{\text{stat}}(0.29)_{\text{sys}} -i22.36(3.63)_{\text{stat}}(0.47)_{\text{sys}}] \mathrm{MeV}$, which is in excellent
agreement with the values in the PDG. The dependence of central mass of $K^*(892)$ on light quark masses are also
evaluated, yielding a consistent  result with previous study.


Our next task in the future is to investigate the evolution of the
broad $\kappa$ resonance as the quark masses vary, which is more
complicated than the $K^*(892)$ .
According to previous lattice results, the pole position of $\kappa$
resonance is unstable in the analysis using the $K$-matrix
parameterization~\cite{Wilson:2019wfr}. This instability may be partly
due to the neglect of left-hand-cut contribution in the simple
$K$-matrix parameterization. We expect the product representation
parameterization would be better than the simple $K$-matrix
analysis. The parameterization based on the unitarized chiral
perturbation theory is also expected to yield stable
results. Furthermore, additional lattice data could provide inputs for
the Roy-Steiner equations to produce more reliable
results.  

\section*{Acknowledgements}
Q.Z. Li would like to think Zheng-Li Wang for providing the $\eta$
meson masess.  We thank the CLQCD collaborations for providing us
their gauge configurations with dynamical fermions, which are
generated on HPC Cluster of ITP-CAS, the Southern Nuclear Science
Computing Center(SNSC), the Siyuan-1 cluster supported by the Center
for High Performance Computing at Shanghai Jiao Tong University and
the Dongjiang Yuan Intelligent Computing Center. This work is
supported by the National Natural Science Foundation of China under
Grant Nos.12335002, 12375078, 12293060, 12293061, 12293063, 12293062, 12525504, 12435002 and 12447101 and Sichuan
Provincial Natural Science Foundation Youth Fund Project
(2026NSFSC0759) . This work is also supported in part by ``The
Fundamental Research Funds for the Central Universities''.
\label{sec:Conclusions}
\newpage
\appendix
\section{Operators}\label{ops}
In this appendix, we list all operators used in the analysis. The
symbol ``$K(p_xp_yp_z)\pi(q_xq_yq_z)$" denotes double-hadron operators
in the square brackets of Eq.~\eqref{DHO} and the single-hadron
$\bar s \gamma_i u$ operators are labeled by ``$K^*_i$" in
Eq.~\eqref{SHO}. We only list the first row for the irrps $E$ and
$T^{-}_1$.
\begin{itemize}
	\item $O_h, \mathbf{d}=(000)$:
	      \begin{flalign*}
		       & O^{T_{1}^-}_0(000) = K^*_{3}[000] &
	      \end{flalign*}

	      \begin{flalign*}
		       & O^{T_{1}^-}_1(000) = K[00-1]\pi[001] -K[001]\pi[00-1] &
	      \end{flalign*}

	      \begin{flalign*}
		                                              & O^{T_{1}^-}_2(000) = K[0-1-1]\pi[011] +K[-10-1]\pi[101] \\ &\qquad\qquad+K[10-1]\pi[-101] +K[01-1]\pi[0-11] -K[-101]\pi[10-1]\\
		                                              & \qquad\qquad-K[101]\pi[-10
		      -1] -K[0-11]\pi[01-1] -K[011]\pi[0-1-1] &
	      \end{flalign*}

	      \begin{flalign*}
		       & O^{T_{1}^-}_3(000) = K[-1-1-1]\pi[111] -K[111]\pi[-1-1-1]              \\&\qquad\qquad+K[1-1-1]\pi[-111] -K[-111]\pi[1-1-1] +K[-11-1]\pi[1-11]\\
		       & \qquad\qquad-K[1-11]\pi[-11-1] -K[-1-11]\pi[11-1] +K[11-1]\pi[-1-11] &
	      \end{flalign*}
	\item $C_{4v}, \mathbf{d}=(001)$:
	      \begin{flalign*}
		       & O^{E}_0(001) = K^*_{2}[001] &
	      \end{flalign*}

	      \begin{flalign*}
		       & O^{E}_1(001) = K[0-11]\pi[010] -K[011]\pi[0-10] &
	      \end{flalign*}

	      \begin{flalign*}
		       & O^{E}_2(001) = -K[010]\pi[0-11] +K[0-10]\pi[011] &
	      \end{flalign*}

	      \begin{flalign*}
		       & O^{E}_3(001) = -K[-1-11]\pi[110] +K[-111]\pi[1-10] -K[1-11]\pi[-110] +K[111]\pi[-1-10] &
	      \end{flalign*}

	      \begin{flalign*}
		       & O^{E}_4(001) = K[110]\pi[-1-11] -K[1-10]\pi[-111] +K[-110]\pi[1-11] -K[-1-10]\pi[111] &
	      \end{flalign*}
	\item $C_{4v}, \mathbf{d}=(002)$:
	      \begin{flalign*}
		       & O^{E}_0(002) = K^*_{2}[002] &
	      \end{flalign*}

	      \begin{flalign*}
		       & O^{E}_1(002) = K[0-11]\pi[011] -K[011]\pi[0-11] &
	      \end{flalign*}

	      \begin{flalign*}
		       & O^{E}_2(002) = -K[-1-11]\pi[111] +K[-111]\pi[1-11] -K[1-11]\pi[-111] +K[111]\pi[-1-11] &
	      \end{flalign*}
	\item $C_{2v}, \mathbf{d}=(011)$:
	      \begin{flalign*}
		       & O^{B_{1}}_0(011) = K^*_{2}[011]-K^*_{3}[011] &
	      \end{flalign*}

	      \begin{flalign*}
		       & O^{B_{1}}_1(011) = K[001]\pi[010] -K[010]\pi[001] &
	      \end{flalign*}

	      \begin{flalign*}
		       & O^{B_{1}}_2(011) = K[-101]\pi[110] -K[-110]\pi[101] +K[101]\pi[-110] -K[110]\pi[-101] &
	      \end{flalign*}

	      \begin{flalign*}
		       & O^{B_{1}}_3(011) = K[020]\pi[0-11] -K[002]\pi[01-1] &
	      \end{flalign*}

	      \begin{flalign*}
		       & O^{B_{1}}_4(011) = K[0-11]\pi[020] -K[01-1]\pi[002] &
	      \end{flalign*}

	      \begin{flalign*}
		       & O^{B_{1}}_5(011) = K[0-10]\pi[021] -K[00-1]\pi[012] &
	      \end{flalign*}

	      \begin{flalign*}
		       & O^{B_{1}}_6(011) = K[012]\pi[00-1] -K[021]\pi[0-10] &
	      \end{flalign*}

	      \begin{flalign*}
		       & O^{B_{2}}_0(011) = K^*_{1}[011] &
	      \end{flalign*}

	      \begin{flalign*}
		       & O^{B_{2}}_1(011) = K[-111]\pi[100] -K[111]\pi[-100] &
	      \end{flalign*}

	      \begin{flalign*}
		       & O^{B_{2}}_2(011) = K[-101]\pi[110] +K[-110]\pi[101] -K[101]\pi[-110] -K[110]\pi[-101] &
	      \end{flalign*}

	      \begin{flalign*}
		       & O^{B_{2}}_3(011) = K[-100]\pi[111] -K[100]\pi[-111] &
	      \end{flalign*}
	\item $C_{3v}, \mathbf{d}=(111)$:
	      \begin{flalign*}
		       & O^{E}_0(111) = K^*_{1}[111]-K^*_{2}[111] &
	      \end{flalign*}

	      \begin{flalign*}
		       & O^{E}_1(111) = -K[011]\pi[100] +K[101]\pi[010] &
	      \end{flalign*}

	      \begin{flalign*}
		       & O^{E}_2(111) = K[100]\pi[011] -K[010]\pi[101] &
	      \end{flalign*}

	      \begin{flalign*}
		       & O^{E}_3(111) = K[200]\pi[-111] -K[020]\pi[1-11] &
	      \end{flalign*}

	      \begin{flalign*}
		       & O^{E}_4(111) = K[-111]\pi[200] -K[1-11]\pi[020] &
	      \end{flalign*}

	      \begin{flalign*}
		       & O^{E}_5(111) = K[121]\pi[0-10] -K[211]\pi[-100] &
	      \end{flalign*}

	      \begin{flalign*}
		       & O^{E}_6(111) = K[-100]\pi[211] -K[0-10]\pi[121] &
	      \end{flalign*}
\end{itemize}

\section{Finite-volume spectra}\label{FVS}
In the following, we list all finite-volume spectra used in the
fitting. The numbers in braces denote the    statistical and systematical
uncertainties in the second column, respectively. The last column
lists the chosen $t_0$ (reference time for GEVP) and $t_e$ for each
energy level.

\begin{multicols}{2}
	\begin{table}[H]
		\centering \renewcommand{\arraystretch}{1.5}
		\begin{tabular}{|c|c|c|}
			\hline
			(   $\boldsymbol d, \Lambda$ ) & $aE_n^{\Lambda, \boldsymbol P}$ & $(t_0, t_e)$ \\
			\hline
			($000$,$T_{1}^-$)              & 0.33475(57)(68)                 & (4, 32)      \\
			\hline
			($000$,$T_{1}^-$)              & 0.39233(48)(190)                & (4, 30)      \\
			\hline
			($001$,$E$)                    & 0.3344(11)(32)                  & (4, 22)      \\
			\hline
			($001$,$E$)                    & 0.40484(54)(51)                 & (4, 26)      \\
			\hline
			($011$,$B_1$)                  & 0.32710(46)(36)                 & (4, 23)      \\
			\hline
			($011$,$B_1$)                  & 0.35481(63)(74)                 & (4, 20)      \\
			\hline
			($011$,$B_2$)                  & 0.3411(15)(22)                  & (4, 16)      \\
			\hline
			($111$,$E$)                    & 0.33238(99)(190)                & (4, 18)      \\
			\hline
			($111$,$E$)                    & 0.36156(51)(66)                 & (4, 22)      \\
			\hline
			($111$,$E$)                    & 0.37769(58)(130)                & (4, 21)      \\
			\hline
			($002$,$E$)                    & 0.3376(21)(55)                  & (4, 16)      \\
			\hline
			($002$,$E$)                    & 0.39605(70)(240)                & (4, 20)      \\
			\hline
		\end{tabular}
		\caption{F48P21}
	\end{table}
	\begin{table}[H]
		\centering

		\renewcommand{\arraystretch}{1.5}
		\begin{tabular}{|c|c|c|}
			\hline
			(   $\boldsymbol d, \Lambda$ ) & $aE_n^{\Lambda, \boldsymbol P}$ & $(t_0, t_e)$ \\
			\hline
			($000$,$T_{1}^-$)              & 0.34209(73)(310)                & (5, 22)      \\
			\hline
			($001$,$E$)                    & 0.3501(16)(98)                  & (4, 20)      \\
			\hline
			($011$,$B_1$)                  & 0.3503(20)(39)                  & (4, 16)      \\
			\hline
			($011$,$B_2$)                  & 0.3654(36)(61)                  & (4, 15)      \\
			\hline
			($111$,$E$)                    & 0.3560(45)(130)                 & (4, 15)      \\
			\hline
		\end{tabular}
		\caption{F32P21}
	\end{table}

	\begin{table}[H]
		\centering \renewcommand{\arraystretch}{1.5}
		\begin{tabular}{|c|c|c|}
			\hline
			($000$,$T_{1}^-$) & 0.45657(38)(30)  & (4, 25) \\
			\hline
			($001$,$E$)       & 0.45549(79)(100) & (4, 20) \\
			\hline
			($011$,$B_1$)     & 0.4482(29)(110)  & (4, 16) \\
			\hline
			($011$,$B_1$)     & 0.51893(98)(140) & (4, 22) \\
			\hline
			($011$,$B_2$)     & 0.4544(14)(40)   & (4, 17) \\
			\hline
			($111$,$E$)       & 0.427(19)(19)    & (4, 16) \\
			\hline
			($111$,$E$)       & 0.5375(14)(38)   & (4, 21) \\
			\hline
			($111$,$E$)       & 0.5557(30)(42)   & (4, 16) \\
			\hline
		\end{tabular}
		\caption{C32P29}
	\end{table}
	\begin{table}[H]
		\centering \renewcommand{\arraystretch}{1.5}
		\begin{tabular}{|c|c|c|}
			\hline
			(   $\boldsymbol d, \Lambda$ ) & $aE_n^{\Lambda, \boldsymbol P}$ & $(t_0, t_e)$ \\
			\hline
			($000$,$T_{1}^-$)              & 0.35225(36)(20)                 & (4, 35)      \\
			\hline
			($000$,$T_{1}^-$)              & 0.42348(38)(70)                 & (4, 35)      \\
			\hline
			($001$,$E$)                    & 0.35275(43)(90)                 & (4, 27)      \\
			\hline
			($001$,$E$)                    & 0.43546(65)(110)                & (4, 29)      \\
			\hline
			($011$,$B_1$)                  & 0.35045(61)(160)                & (4, 23)      \\
			\hline
			($011$,$B_1$)                  & 0.38301(52)(150)                & (4, 27)      \\
			\hline
			($011$,$B_2$)                  & 0.35192(93)(160)                & (4, 21)      \\
			\hline
			($111$,$E$)                    & 0.35321(65)(240)                & (4, 19)      \\
			\hline
			($111$,$E$)                    & 0.39517(38)(40)                 & (4, 26)      \\
			\hline
			($111$,$E$)                    & 0.40627(84)(170)                & (4, 23)      \\
			\hline
			($002$,$E$)                    & 0.3570(13)(44)                  & (4, 18)      \\
			\hline
			($002$,$E$)                    & 0.41888(95)(380)                & (4, 27)      \\
			\hline
		\end{tabular}
		\caption{F48P30}
	\end{table}
	\begin{table}[H]
		\centering \renewcommand{\arraystretch}{1.5}
		\begin{tabular}{|c|c|c|}
			\hline
			(   $\boldsymbol d, \Lambda$ ) & $aE_n^{\Lambda, \boldsymbol P}$ & $(t_0, t_e)$ \\
			\hline
			($000$,$T_{1}^-$)              & 0.35449(56)(38)                 & (4, 33)      \\
			\hline
			($001$,$E$)                    & 0.35830(81)(300)                & (4, 22)      \\
			\hline
			($011$,$B_1$)                  & 0.3524(17)(26)                  & (4, 18)      \\
			\hline
			($011$,$B_2$)                  & 0.3598(18)(120)                 & (4, 18)      \\
			\hline
			($111$,$E$)                    & 0.3592(38)(140)                 & (4, 16)      \\
			\hline
		\end{tabular}
		\caption{F32P30}
	\end{table}
	\begin{table}[H]
		\centering

		\renewcommand{\arraystretch}{1.5}
		\begin{tabular}{|c|c|c|}
			\hline
			(   $\boldsymbol d, \Lambda$ ) & $aE_n^{\Lambda, \boldsymbol P}$ & $(t_0, t_e)$ \\
			\hline
			($000$,$T_{1}^-$)              & 0.24895(32)(44)                 & (10, 44)     \\
			\hline
			($001$,$E$)                    & 0.24911(86)(120)                & (10, 35)     \\
			\hline
			($011$,$B_1$)                  & 0.24549(120)(96)                & (10, 28)     \\
			\hline
			($011$,$B_2$)                  & 0.2501(25)(67)                  & (10, 24)     \\
			\hline
			($111$,$E$)                    & 0.2385(42)(93)                  & (10, 23)     \\
			\hline
			($002$,$E$)                    & 0.2519(28)(82)                  & (10, 21)     \\
			\hline
		\end{tabular}
		\caption{H48P32}
	\end{table}

	\begin{table}[H]
		\centering

		\renewcommand{\arraystretch}{1.5}
		\begin{tabular}{|c|c|c|}
			\hline
			(   $\boldsymbol d, \Lambda$ ) & $aE_n^{\Lambda, \boldsymbol P}$ & $(t_0, t_e)$ \\
			\hline
			($000$,$T_{1}^-$)              & 0.43122(67)(47)                 & (4, 28)      \\
			\hline
			($000$,$T_{1}^-$)              & 0.46091(54)(79)                 & (4, 26)      \\
			\hline
			($000$,$T_{1}^-$)              & 0.53101(65)(110)                & (4, 23)      \\
			\hline
			($001$,$E$)                    & 0.4330(11)(35)                  & (4, 22)      \\
			\hline
			($001$,$E$)                    & 0.46473(63)(110)                & (4, 24)      \\
			\hline
			($001$,$E$)                    & 0.48651(50)(49)                 & (4, 23)      \\
			\hline
			($001$,$E$)                    & 0.53719(64)(52)                 & (4, 21)      \\
			\hline
			($001$,$E$)                    & 0.5509(10)(19)                  & (4, 20)      \\
			\hline
			($011$,$B_1$)                  & 0.40673(61)(52)                 & (4, 27)      \\
			\hline
			($011$,$B_1$)                  & 0.44582(89)(160)                & (4, 17)      \\
			\hline
			($011$,$B_1$)                  & 0.49573(60)(61)                 & (4, 21)      \\
			\hline
			($011$,$B_1$)                  & 0.5124(20)(57)                  & (4, 20)      \\
			\hline
			($011$,$B_1$)                  & 0.5424(15)(51)                  & (4, 19)      \\
			\hline
			($011$,$B_2$)                  & 0.4360(13)(51)                  & (4, 17)      \\
			\hline
			($011$,$B_2$)                  & 0.4682(16)(30)                  & (4, 21)      \\
			\hline
			($011$,$B_2$)                  & 0.4950(15)(47)                  & (4, 20)      \\
			\hline
			($011$,$B_2$)                  & 0.5044(25)(42)                  & (4, 19)      \\
			\hline
			($111$,$E$)                    & 0.4122(14)(35)                  & (4, 25)      \\
			\hline
			($111$,$E$)                    & 0.43684(98)(240)                & (4, 17)      \\
			\hline
			($111$,$E$)                    & 0.4522(33)(96)                  & (4, 15)      \\
			\hline
			($111$,$E$)                    & 0.5176(15)(24)                  & (4, 18)      \\
			\hline
		\end{tabular}
		\caption{C48P14}

	\end{table}
	\begin{table}[H]
		\centering

		\renewcommand{\arraystretch}{1.5}
		\begin{tabular}{|c|c|c|}
			\hline
			(   $\boldsymbol d, \Lambda$ ) & $aE_n^{\Lambda, \boldsymbol P}$ & $(t_0, t_e)$ \\
			\hline
			($000$,$t_{1}^-$)              & 0.43627(38)(23)                 & (4, 30)      \\
			\hline
			($000$,$t_{1}^-$)              & 0.47175(33)(11)                 & (4, 30)      \\
			\hline
			($001$,$e$)                    & 0.43717(46)(19)                 & (4, 22)      \\
			\hline
			($001$,$e$)                    & 0.480010(370)(80)               & (4, 28)      \\
			\hline
			($001$,$e$)                    & 0.49476(39)(14)                 & (4, 27)      \\
			\hline
			($011$,$b_1$)                  & 0.42267(40)(45)                 & (8, 26)      \\
			\hline
			($011$,$b_1$)                  & 0.4436(15)(39)                  & (8, 21)      \\
			\hline
			($011$,$b_1$)                  & 0.50537(54)(76)                 & (8, 24)      \\
			\hline
			($011$,$b_2$)                  & 0.4331(11)(59)                  & (4, 18)      \\
			\hline
			($011$,$b_2$)                  & 0.48805(41)(16)                 & (4, 26)      \\
			\hline
			($011$,$b_2$)                  & 0.50698(41)(21)                 & (4, 24)      \\
			\hline
			($011$,$b_2$)                  & 0.51254(90)(320)                & (4, 22)      \\
			\hline
			($111$,$e$)                    & 0.42786(71)(240)                & (4, 18)      \\
			\hline
			($111$,$e$)                    & 0.44365(59)(680)                & (4, 22)      \\
			\hline
			($111$,$e$)                    & 0.45856(50)(140)                & (4, 21)      \\
			\hline
			($002$,$e$)                    & 0.43654(83)(110)                & (4, 18)      \\
			\hline
			($002$,$e$)                    & 0.47277(52)(85)                 & (4, 20)      \\
			\hline
		\end{tabular}
		\caption{C48P23}
	\end{table}
\end{multicols}
\section{Energy levels comparison}\label{app:compare}

\begin{figure}[H]
	\centering
	\begin{subfigure}{1\textwidth}
		\centering
		\includegraphics[width=\textwidth]{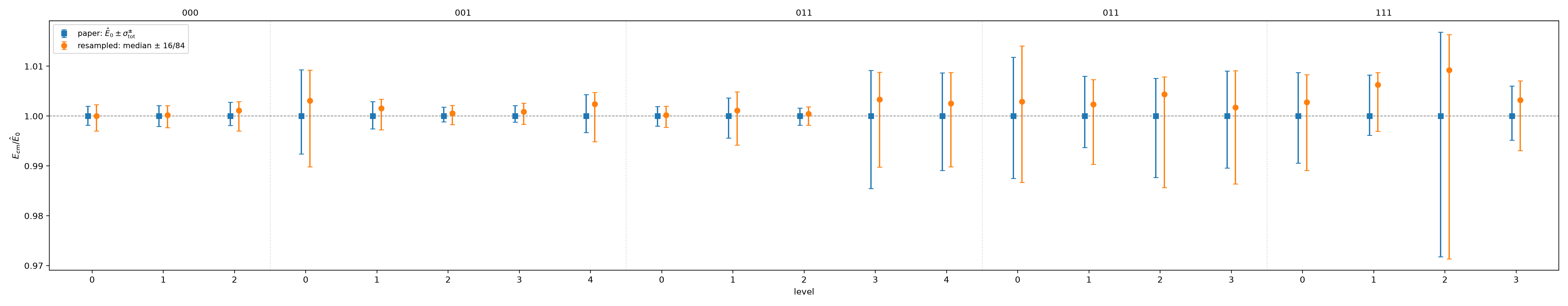}
		\caption{C48P14 }
	\end{subfigure}
	\hfill
	\begin{subfigure}{1\textwidth}
		\centering
		\includegraphics[width=\textwidth]{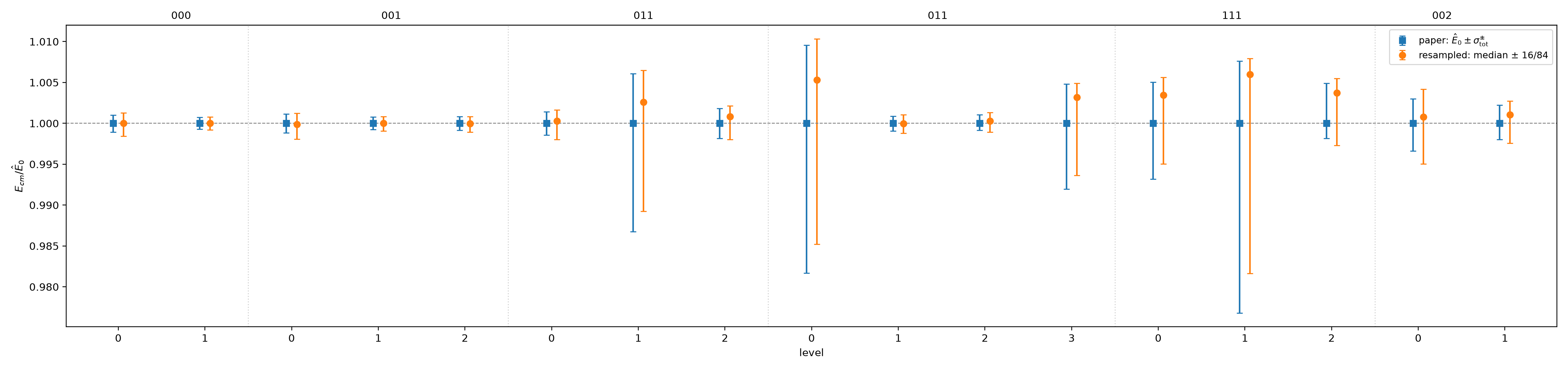}
		\caption{C48P23 }
	\end{subfigure}
	\hfill
	\begin{subfigure}{0.45\textwidth}
		\centering
		\includegraphics[width=\textwidth]{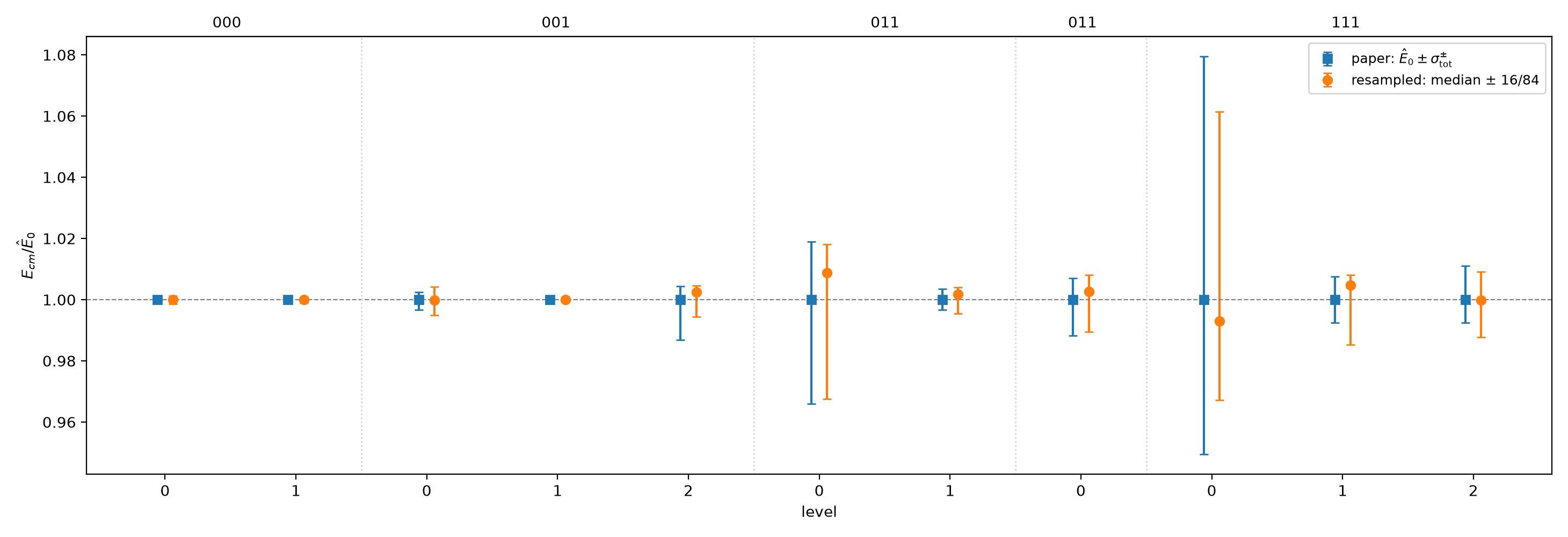}
		\caption{C32P29 }
	\end{subfigure}
	\hfill
	\begin{subfigure}{0.45\textwidth}
		\centering
		\includegraphics[width=\textwidth]{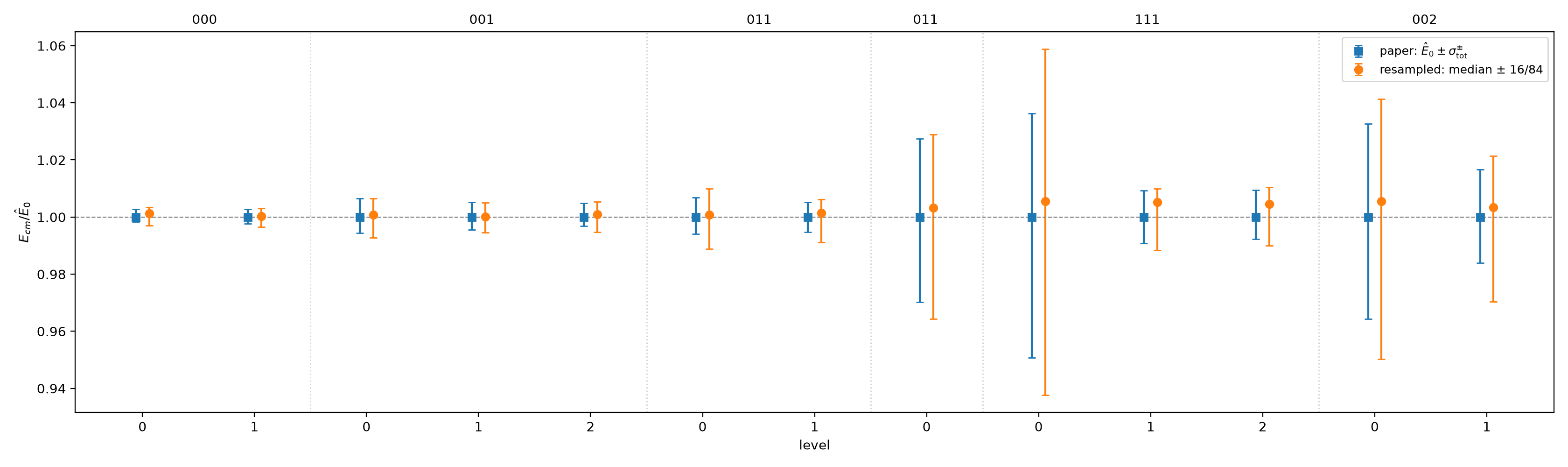}
		\caption{H48P32}
	\end{subfigure}
	\hfill
	\begin{subfigure}{0.45\textwidth}
		\centering
		\includegraphics[width=\textwidth]{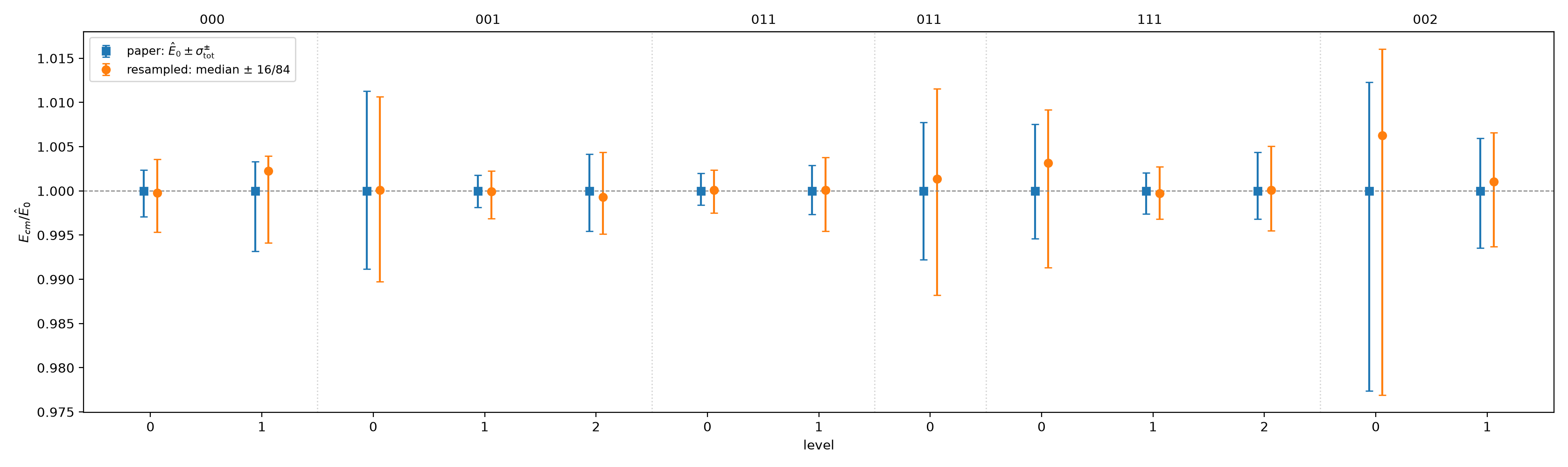}
		\caption{F48P21}
	\end{subfigure}
	\hfill
	\begin{subfigure}{0.45\textwidth}
		\centering
		\includegraphics[width=\textwidth]{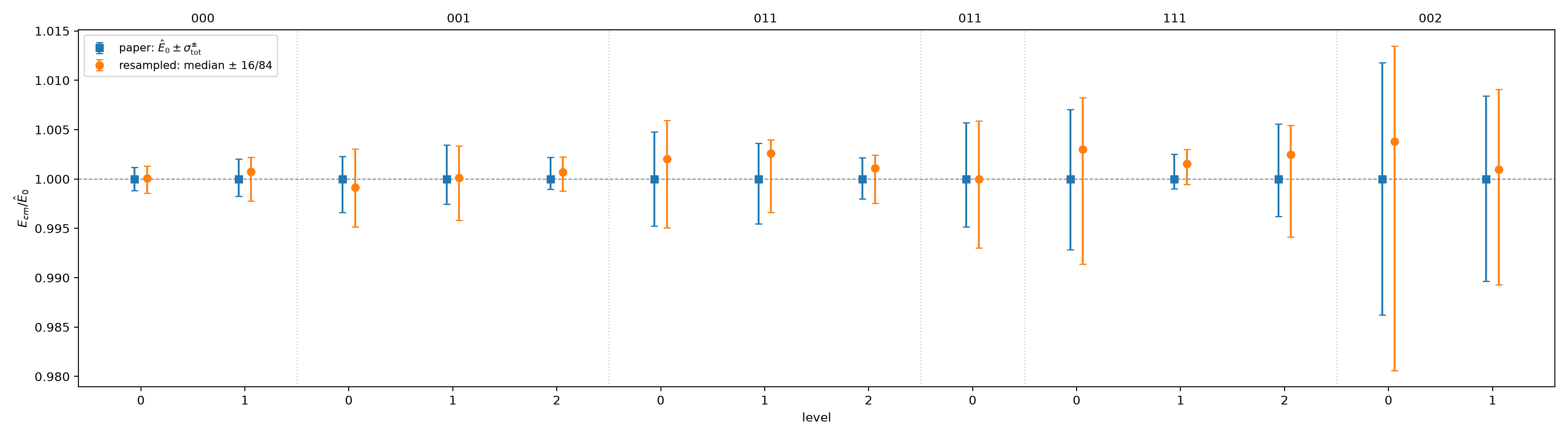}
		\caption{F48P30}
	\end{subfigure}

	\vspace{1ex}
	\begin{subfigure}{0.45\textwidth}
		\centering
		\includegraphics[width=\textwidth]{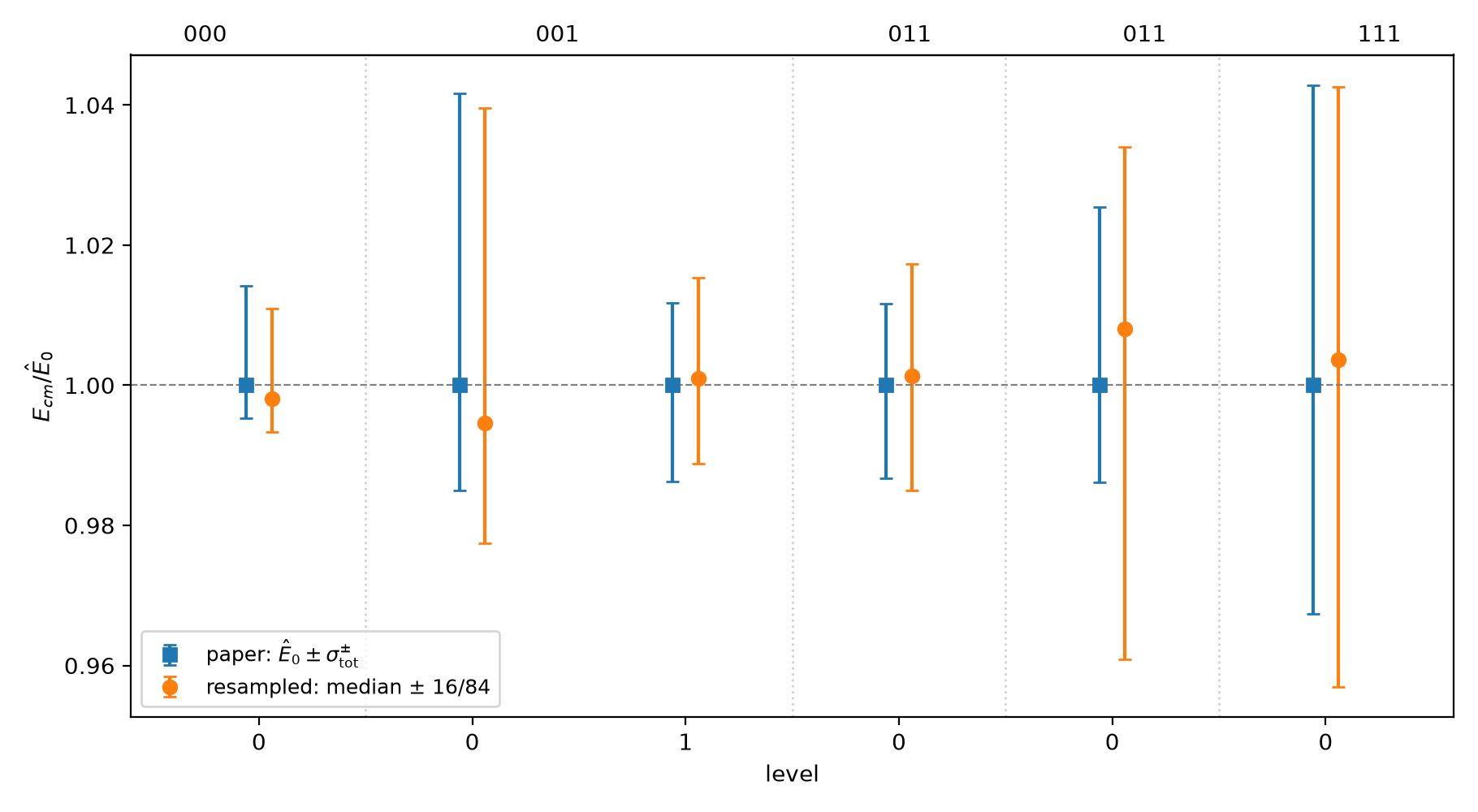}
		\caption{F32P21}
	\end{subfigure}
	\hfill
	\begin{subfigure}{0.45\textwidth}
		\centering
		\includegraphics[width=\textwidth]{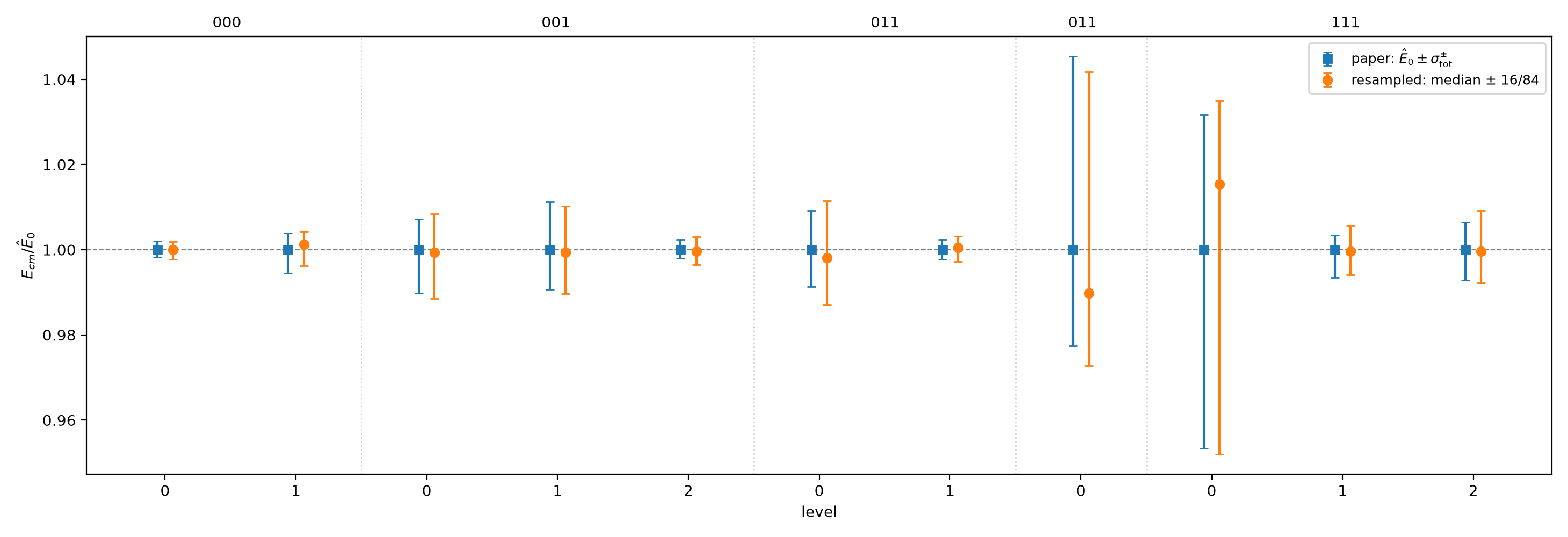}
		\caption{F32P30}
	\end{subfigure}

	\caption{{ Comparison of the finite-volume energy levels obtained
				directly from the AIC-weighted average,
				Eqs.~\eqref{eq:sys}--\eqref{eq:totay} (blue squares, $E_1$), and from
				the AIC-weighted resampling of the fit windows that serves as input
				to the L\"uscher analysis (orange circles, $E_2$), for all ensembles
				used in this work.  All energies are boosted to the
				centre-of-mass frame and normalized to $E_1$, so that the blue points
				lie at unity by construction (gray dashed line). }
	}
	\label{fig:resample_compare}
\end{figure}

{
\section{Cross-check of uncertainty propagation on C48P14}\label{app:C48P14}

In order to check the influence of correlations, a full propagation of uncertainties is performed on ensemble C48P14 using the BW parameterization. Specifically, 300 fit windows are drawn with probability given by the corresponding AIC weight. For each fit window, the energy-extraction fitting is performed on 2000 bootstrap samples. Finally, we perform a total of $6\times10^{6}$ L\"uscher analyses to propagate the statistical and systematic uncertainties, where each L\"uscher analysis needs $1\sim 15$ minutes.The fitting parameters and the pole position corresponding to the $K^*(892)$ resonance turn out to be consistent with those in the manuscript, as shown in Table~\ref{tab:C48P14-crosscheck}.
\begin{table}[H]
	\centering
	\vspace{4mm}
\renewcommand{\arraystretch}{1.5} 
	  \begin{tabular}{cccc}
    \toprule
     & $g_1^2$ & $m_1^2$ & $\sqrt{s_0}$ [MeV] \\
    \midrule
    a)
     & $0.178(9)_{\rm stat}\binom{+0.01}{-0.017}_{\rm sys}$
      & $0.1946(6)_{\rm stat}(15)_{\rm sys}$
      & $825.3(1.1)_{\rm stat}(3.0)_{\rm sys}
         -i\,12.8(0.7)_{\rm stat}\binom{+1.0}{-1.5}_{\rm sys}$ \\
   b)
          & $0.181^{+0.021}_{-0.025}$
      & $0.1931^{+0.0018}_{-0.0022}$
      & $822.4^{+3.8}_{-4.6}-i\,12.7^{+1.4}_{-1.8}$ \\
    \bottomrule
  \end{tabular}
  \caption{Comparison of the uncertainty propagation for C48P14 using the
  BW parameterization.}
	\label{tab:C48P14-crosscheck}
\end{table}
The central value, statistical and systematic uncertainties labeled by II are evaluated as Ref.~\cite{Boyle:2024grr, Boyle:2024hvv}:
\begin{itemize}
	\item Central value:
	      \begin{equation}
		      \label{eq:cv-crosscheck}
		      \hat p_{b =0 } = \sum_{k=1}^{300}\omega_{k}p_{b=0},\quad p=g_{1}\ \text{or}\  m_{1}\ ,
	      \end{equation}
	      where $b=0$ indicates the gauge-average result.
	\item Statistical uncertainty:
	      \begin{equation}
		      \label{eq:su-crosscheck}
		      \sigma_{stat} = \sqrt{\frac{1}{N_{boot} - 1}\sum_{b=1}^{N_{boot}}(\hat p_{b} - \bar{\hat p})^2}, \quad \bar{\hat p} = \frac{1}{N_{boot}}\sum_{b=1}^{N_{boot}} \hat p_{b}
	      \end{equation}
	\item Systematic uncertainty:
	      \begin{equation}
		      \label{eq:sysu-crosscheck}
          \sigma^\pm_{sys} = p_{sys\pm} - \hat p_{b=0} \ ,
	      \end{equation}
	      where $p_{sys\pm}$ are respectively the 84th and 16th $\omega-$weighted percentiles of $p_{b=0}$.
\end{itemize}

We spent at least $ 10^{5}$ CPU hours to obtain these results, and that is only for a single ensemble and a single parameterization. Thus, the method adopted in the manuscript is a compromise, given the limitations on our available computational resources.

}

\newpage
\bibliographystyle{h-physrev} \bibliography{ref}

@article{Liu:2026gxr,
    author = "Liu, Hang and Liu, Liuming and Tan, Jin-Xin and Wang, Wei and Yan, Haobo and Zhu, Qian-Teng",
    title = "{A Lattice QCD study of $p-Λ$ scattering in continuum and chiral limits}",
    eprint = "2603.05854",
    archivePrefix = "arXiv",
    primaryClass = "hep-lat",
    month = "3",
    year = "2026"
}

@article{Xiao:2000kx,
    author = "Xiao, Zhiguang and Zheng, H. Q.",
    title = "{Left-hand singularities, hadron form-factors and the properties of the sigma meson}",
    eprint = "hep-ph/0011260",
    archivePrefix = "arXiv",
    doi = "10.1016/S0375-9474(01)01100-9",
    journal = "Nucl. Phys. A",
    volume = "695",
    pages = "273--294",
    year = "2001"
}

@article{CLQCD:2023sdb,
    author = "Hu, Zhi-Cheng and others",
    collaboration = "CLQCD",
    title = "{Quark masses and low-energy constants in the continuum from the tadpole-improved clover ensembles}",
    eprint = "2310.00814",
    archivePrefix = "arXiv",
    primaryClass = "hep-lat",
    doi = "10.1103/PhysRevD.109.054507",
    journal = "Phys. Rev. D",
    volume = "109",
    number = "5",
    pages = "054507",
    year = "2024"
}

@article{CLQCD:2024yyn,
    author = "Du, Hai-Yang and others",
    collaboration = "CLQCD",
    title = "{Charmed meson masses and decay constants in the continuum limit from the tadpole improved clover ensembles}",
    eprint = "2408.03548",
    archivePrefix = "arXiv",
    primaryClass = "hep-lat",
    doi = "10.1103/PhysRevD.111.054504",
    journal = "Phys. Rev. D",
    volume = "111",
    number = "5",
    pages = "054504",
    year = "2025"
}

@article{ParticleDataGroup:2024cfk,
  author = "Navas, S. and others",
  collaboration = "Particle Data Group",
  title = "{Review of particle physics}",
  doi = "10.1103/PhysRevD.110.030001",
  journal = "Phys. Rev. D",
  volume = "110",
  number = "3",
  pages = "030001",
  year = "2024",
}

@article{Luscher:1990ux,
  author = "Luscher, Martin",
  title = "{Two particle states on a torus and their relation to the scattering
           matrix}",
  reportNumber = "DESY-90-131",
  doi = "10.1016/0550-3213(91)90366-6",
  journal = "Nucl. Phys. B",
  volume = "354",
  pages = "531--578",
  year = "1991",
}

@article{Rummukainen:1995vs,
  author = "Rummukainen, K. and Gottlieb, Steven A.",
  title = "{Resonance scattering phase shifts on a nonrest frame lattice}",
  eprint = "hep-lat/9503028",
  archivePrefix = "arXiv",
  reportNumber = "IUHET-297",
  doi = "10.1016/0550-3213(95)00313-H",
  journal = "Nucl. Phys. B",
  volume = "450",
  pages = "397--436",
  year = "1995",
}

@article{Frazer:1961zz,
  author = "Frazer, William R.",
  title = "{Applications of Conformal Mapping to the Phenomenological
           Representation of Scattering Amplitudes}",
  doi = "10.1103/PhysRev.123.2180",
  journal = "Phys. Rev.",
  volume = "123",
  pages = "2180--2182",
  year = "1961",
}

@article{Akaike:1974vps,
  author = "Akaike, H.",
  title = "{A new look at the statistical model identification}",
  doi = "10.1109/TAC.1974.1100705",
  journal = "IEEE Trans. Automatic Control",
  volume = "19",
  number = "6",
  pages = "716--723",
  year = "1974",
}

@article{Boyle:2024hvv,
  author = {Boyle, Peter and Erben, Felix and G\"ulpers, Vera and Hansen,
            Maxwell T. and Joswig, Fabian and Marshall, Michael and Lachini,
            Nelson Pitanga and Portelli, Antonin},
  title = "{Light and Strange Vector Resonances from Lattice QCD at Physical
           Quark Masses}",
  eprint = "2406.19194",
  archivePrefix = "arXiv",
  primaryClass = "hep-lat",
  reportNumber = "CERN-TH-2024-088",
  doi = "10.1103/PhysRevLett.134.111901",
  journal = "Phys. Rev. Lett.",
  volume = "134",
  number = "11",
  pages = "111901",
  year = "2025",
}

@article{HadronSpectrum:2009krc,
  author = "Peardon, Michael and Bulava, John and Foley, Justin and Morningstar,
            Colin and Dudek, Jozef and Edwards, Robert G. and Joo, Balint and Lin
            , Huey-Wen and Richards, David G. and Juge, Keisuke Jimmy",
  collaboration = "Hadron Spectrum",
  title = "{A Novel quark-field creation operator construction for hadronic
           physics in lattice QCD}",
  eprint = "0905.2160",
  archivePrefix = "arXiv",
  primaryClass = "hep-lat",
  reportNumber = "JLAB-THY-09-985",
  doi = "10.1103/PhysRevD.80.054506",
  journal = "Phys. Rev. D",
  volume = "80",
  pages = "054506",
  year = "2009",
}

@article{Hite:1973pm,
  author = "Hite, G. E. and Steiner, F.",
  title = "{New dispersion relations and their application to partial-wave
           amplitudes}",
  doi = "10.1007/BF02722827",
  journal = "Nuovo Cim. A",
  volume = "18",
  pages = "237--270",
  year = "1973",
}

@article{Ditsche:2012fv,
  author = "Ditsche, C. and Hoferichter, M. and Kubis, B. and Mei\ss{}ner,
            Ulf-G.",
  title = "{Roy-Steiner equations for pion-nucleon scattering}",
  eprint = "1203.4758",
  archivePrefix = "arXiv",
  primaryClass = "hep-ph",
  doi = "10.1007/JHEP06(2012)043",
  journal = "JHEP",
  volume = "06",
  pages = "043",
  year = "2012",
}

@article{Hoferichter:2015hva,
  author = "Hoferichter, Martin and Ruiz de Elvira, Jacobo and Kubis, Bastian
            and Mei\ss{}ner, Ulf-G.",
  title = "{Roy\textendash{}Steiner-equation analysis of pion\textendash{}
           nucleon scattering}",
  eprint = "1510.06039",
  archivePrefix = "arXiv",
  primaryClass = "hep-ph",
  reportNumber = "INT-PUB-15-050",
  doi = "10.1016/j.physrep.2016.02.002",
  journal = "Phys. Rept.",
  volume = "625",
  pages = "1--88",
  year = "2016",
}

@article{Hoferichter:2023mgy,
  author = "Hoferichter, Martin and de Elvira, Jacobo Ruiz and Kubis, Bastian
            and Mei\ss{}ner, Ulf-G.",
  title = "{Nucleon resonance parameters from Roy\textendash{}Steiner equations}
           ",
  eprint = "2312.15015",
  archivePrefix = "arXiv",
  primaryClass = "hep-ph",
  reportNumber = "IPARCOS-UCM-23-139",
  doi = "10.1016/j.physletb.2024.138698",
  journal = "Phys. Lett. B",
  volume = "853",
  pages = "138698",
  year = "2024",
}

@article{Cao:2023ntr,
  author = "Cao, Xiong-Hui and Li, Qu-Zhi and Guo, Zhi-Hui and Zheng, Han-Qing",
  title = "{Roy equation analyses of {\ensuremath{\pi}}{\ensuremath{\pi}}
           scatterings at unphysical pion masses}",
  eprint = "2303.02596",
  archivePrefix = "arXiv",
  primaryClass = "hep-ph",
  doi = "10.1103/PhysRevD.108.034009",
  journal = "Phys. Rev. D",
  volume = "108",
  number = "3",
  pages = "034009",
  year = "2023",
}

@article{Cao:2024zuy,
  author = "Cao, Xiong-Hui and Guo, Feng-Kun and Guo, Zhi-Hui and Li, Qu-Zhi",
  title = "{Rigorous Roy-Steiner equation analysis of {\ensuremath{\pi}}K
           scattering at unphysical quark masses}",
  eprint = "2412.03374",
  archivePrefix = "arXiv",
  primaryClass = "hep-ph",
  doi = "10.1103/38b9-m57d",
  journal = "Phys. Rev. D",
  volume = "112",
  number = "3",
  pages = "L031503",
  year = "2025",
}

@article{Lyu:2024lzr,
  author = "Lyu, Yuan-Lin and Li, Qu-Zhi and Xiao, Zhiguang and Zheng, Han-Qing",
  title = "{Revisiting O(N) {\ensuremath{\sigma}} model at unphysical pion
           masses and high temperatures}",
  eprint = "2402.19243",
  archivePrefix = "arXiv",
  primaryClass = "hep-ph",
  doi = "10.1103/PhysRevD.109.094026",
  journal = "Phys. Rev. D",
  volume = "109",
  number = "9",
  pages = "094026",
  year = "2024",
}

@article{Li:2025fvg,
  author = "Li, Qu-Zhi and Xiao, Zhiguang and Zheng, Han-Qing",
  title = "{On the pole trajectory of the subthreshold negative parity nucleon
           with varying pion masses*}",
  eprint = "2501.01619",
  archivePrefix = "arXiv",
  primaryClass = "hep-ph",
  doi = "10.1088/1674-1137/adfa02",
  journal = "Chin. Phys.",
  volume = "49",
  number = "12",
  pages = "123103",
  year = "2025",
}

@article{Rodas:2023nec,
  author = "Rodas, Arkaitz and Dudek, Jozef J. and Edwards, Robert G.",
  collaboration = "Hadron Spectrum",
  title = "{Determination of crossing-symmetric {\ensuremath{\pi}}{\ensuremath{
           \pi}} scattering amplitudes and the quark mass evolution of the {
           \ensuremath{\sigma}} constrained by lattice QCD}",
  eprint = "2304.03762",
  archivePrefix = "arXiv",
  primaryClass = "hep-lat",
  reportNumber = "JLAB-THY-23-3791",
  doi = "10.1103/PhysRevD.109.034513",
  journal = "Phys. Rev. D",
  volume = "109",
  number = "3",
  pages = "034513",
  year = "2024",
}

@article{Wilson:2014cna,
  author = "Wilson, David J. and Dudek, Jozef J. and Edwards, Robert G. and
            Thomas, Christopher E.",
  title = "{Resonances in coupled $\pi K, \eta K$ scattering from lattice QCD}",
  eprint = "1411.2004",
  archivePrefix = "arXiv",
  primaryClass = "hep-ph",
  reportNumber = "JLAB-THY-14-1982, DAMTP-2014-81, JLAB-THY-14-1892",
  doi = "10.1103/PhysRevD.91.054008",
  journal = "Phys. Rev. D",
  volume = "91",
  number = "5",
  pages = "054008",
  year = "2015",
}

@article{Briceno:2016mjc,
  author = "Briceno, Raul A. and Dudek, Jozef J. and Edwards, Robert G. and
            Wilson, David J.",
  title = "{Isoscalar $\pi\pi$ scattering and the $\sigma$ meson resonance from
           QCD}",
  eprint = "1607.05900",
  archivePrefix = "arXiv",
  primaryClass = "hep-ph",
  reportNumber = "JLAB-THY-16-2287, DAMTP-2016-47",
  doi = "10.1103/PhysRevLett.118.022002",
  journal = "Phys. Rev. Lett.",
  volume = "118",
  number = "2",
  pages = "022002",
  year = "2017",
}

@article{Cao:2022zhn,
  author = "Cao, Xiong-Hui and Li, Qu-Zhi and Zheng, Han-Qing",
  title = "{A possible subthreshold pole in S$_{11}$ channel from \ensuremath{
           \pi}N Roy-Steiner equation analyses}",
  eprint = "2207.09743",
  archivePrefix = "arXiv",
  primaryClass = "hep-ph",
  doi = "10.1007/JHEP12(2022)073",
  journal = "JHEP",
  volume = "12",
  pages = "073",
  year = "2022",
}

@article{Caprini:2005zr,
  author = "Caprini, Irinel and Colangelo, Gilberto and Leutwyler, Heinrich",
  title = "{Mass and width of the lowest resonance in QCD}",
  eprint = "hep-ph/0512364",
  archivePrefix = "arXiv",
  doi = "10.1103/PhysRevLett.96.132001",
  journal = "Phys. Rev. Lett.",
  volume = "96",
  pages = "132001",
  year = "2006",
}

@article{Roy:1971tc,
  author = "Roy, S. M.",
  title = "{Exact integral equation for pion pion scattering involving only
           physical region partial waves}",
  doi = "10.1016/0370-2693(71)90724-6",
  journal = "Phys. Lett. B",
  volume = "36",
  pages = "353--356",
  year = "1971",
}

@article{Okubo:1961jc,
  author = "Okubo, Susumu",
  title = "{Note on unitary symmetry in strong interactions}",
  doi = "10.1143/PTP.27.949",
  journal = "Prog. Theor. Phys.",
  volume = "27",
  pages = "949--966",
  year = "1962",
}

@article{Wilson:2019wfr,
  author = "Wilson, David J. and Briceno, Raul A. and Dudek, Jozef J. and
            Edwards, Robert G. and Thomas, Christopher E.",
  title = "{The quark-mass dependence of elastic $\pi K$ scattering from QCD}",
  eprint = "1904.03188",
  archivePrefix = "arXiv",
  primaryClass = "hep-lat",
  reportNumber = "DAMTP-2019-13, JLAB-THY-19-2911",
  doi = "10.1103/PhysRevLett.123.042002",
  journal = "Phys. Rev. Lett.",
  volume = "123",
  number = "4",
  pages = "042002",
  year = "2019",
}

@article{Helmes:2018nug,
  author = "Helmes, Christopher and Jost, Christian and Knippschild, Bastian and
            Kostrzewa, Bartosz and Liu, Liuming and Pittler, Ferenc and Urbach,
            Carsten and Werner, Markus",
  collaboration = "ETM",
  title = "{Hadron-Hadron Interactions from $N_f=2+1+1$ Lattice QCD: $I=3/2$ $
           \pi K$ Scattering Length}",
  eprint = "1809.08886",
  archivePrefix = "arXiv",
  primaryClass = "hep-lat",
  doi = "10.1103/PhysRevD.98.114511",
  journal = "Phys. Rev. D",
  volume = "98",
  number = "11",
  pages = "114511",
  year = "2018",
}

@article{Beane_2006,
  doi = {10.1103/physrevd.74.114503},
  url = {https://doi.org/10.1103%2Fphysrevd.74.114503},
  year = 2006,
  month = {dec},
  publisher = {American Physical Society ({APS})},
  volume = {74},
  number = {11},
  author = {Silas R. Beane and Paulo F. Bedaque and Thomas C. Luu and Kostas
            Orginos and Elisabetta Pallante and Assumpta Parre{\~{n} }o and
            Martin J. Savage},
  title = {$\less$mml:math xmlns:mml="http://www.w3.org/1998/Math/{MathML}"
           display="inline"$\greater$$\less$mml:mi$\greater$$\uppi$$\less
           $/mml:mi$\greater$$\less$mml:mi$\greater$K$\less$/mml:mi$\greater$$
           \less$/mml:math$\greater$scattering in full {QCD} with domain-wall
           valence quarks},
  journal = {Physical Review D},
}

@article{Fu_2012,
  doi = {10.1103/physrevd.85.074501},
  url = {https://doi.org/10.1103%2Fphysrevd.85.074501},
  year = 2012,
  month = {apr},
  publisher = {American Physical Society ({APS})},
  volume = {85},
  number = {7},
  author = {Ziwen Fu},
  title = {Lattice study on$\less$mml:math xmlns:mml="
           http://www.w3.org/1998/Math/{MathML}" display="inline"$\greater$$\less
           $mml:mi$\greater$$\uppi$$\less$/mml:mi$\greater$$\less$mml:mi$\greater
           $K$\less$/mml:mi$\greater$$\less$/mml:math$\greater$scattering with
           moving wall source},
  journal = {Physical Review D},
}

@article{Lang_2012,
  doi = {10.1103/physrevd.86.054508},
  url = {https://doi.org/10.1103%2Fphysrevd.86.054508},
  year = 2012,
  month = {sep},
  publisher = {American Physical Society ({APS})},
  volume = {86},
  number = {5},
  author = {C. B. Lang and Luka Leskovec and Daniel Mohler and Sasa Prelovsek},
  title = {$\less$mml:math xmlns:mml="http://www.w3.org/1998/Math/{MathML}"
           display="inline"$\greater$$\less$mml:mi$\greater$K$\less$/mml:mi$
           \greater$$\less$/mml:math$\greater$$\less$mml:math xmlns:mml="
           http://www.w3.org/1998/Math/{MathML}" display="inline"$\greater$$\less
           $mml:mi$\greater$$\uppi$$\less$/mml:mi$\greater$$\less$/mml:math$
           \greater$scattering for isospin$\less$mml:math xmlns:mml="
           http://www.w3.org/1998/Math/{MathML}" display="inline"$\greater$$\less
           $mml:mfrac$\greater$$\less$mml:mn$\greater$1$\less$/mml:mn$\greater$$
           \less$mml:mn$\greater$2$\less$/mml:mn$\greater$$\less$/mml:mfrac$
           \greater$$\less$/mml:math$\greater$and$\less$mml:math xmlns:mml="
           http://www.w3.org/1998/Math/{MathML}" display="inline"$\greater$$\less
           $mml:mfrac$\greater$$\less$mml:mn$\greater$3$\less$/mml:mn$\greater$$
           \less$mml:mn$\greater$2$\less$/mml:mn$\greater$$\less$/mml:mfrac$
           \greater$$\less$/mml:math$\greater$in lattice {QCD} },
  journal = {Physical Review D},
}

@article{Dudek:2012gj,
  author = "Dudek, Jozef J. and Edwards, Robert G. and Thomas, Christopher E.",
  title = "{S and D-wave phase shifts in isospin-2 pi pi scattering from lattice
           QCD}",
  eprint = "1203.6041",
  archivePrefix = "arXiv",
  primaryClass = "hep-ph",
  reportNumber = "JLAB-THY-12-1504, TCDMATH-12-03",
  doi = "10.1103/PhysRevD.86.034031",
  journal = "Phys. Rev. D",
  volume = "86",
  pages = "034031",
  year = "2012",
}

@article{Yao:2020bxx,
  author = "Yao, De-Liang and Dai, Ling-Yun and Zheng, Han-Qing and Zhou,
            Zhi-Yong",
  title = "{A review on partial-wave dynamics with chiral effective field theory
           and dispersion relation}",
  eprint = "2009.13495",
  archivePrefix = "arXiv",
  primaryClass = "hep-ph",
  doi = "10.1088/1361-6633/abfa6f",
  journal = "Rept. Prog. Phys.",
  volume = "84",
  number = "7",
  pages = "076201",
  year = "2021",
}

@article{Bernard:2008ax,
  author = "Bernard, Veronique and Lage, Michael and Meissner, Ulf-G. and
            Rusetsky, Akaki",
  title = "{Resonance properties from the finite-volume energy spectrum}",
  eprint = "0806.4495",
  archivePrefix = "arXiv",
  primaryClass = "hep-lat",
  reportNumber = "HISKP-TH-08-08, FZJ-IKP(TH)-08-10",
  doi = "10.1088/1126-6708/2008/08/024",
  journal = "JHEP",
  volume = "08",
  pages = "024",
  year = "2008",
}

@article{Gockeler:2012yj,
  author = "Gockeler, M. and Horsley, R. and Lage, M. and Meissner, U. -G. and
            Rakow, P. E. L. and Rusetsky, A. and Schierholz, G. and Zanotti, J.
            M.",
  title = "{Scattering phases for meson and baryon resonances on general
           moving-frame lattices}",
  eprint = "1206.4141",
  archivePrefix = "arXiv",
  primaryClass = "hep-lat",
  reportNumber = "ADP-12-27-T794, DESY-12-098, EDINBURGH-2012-09, LTH-948",
  doi = "10.1103/PhysRevD.86.094513",
  journal = "Phys. Rev. D",
  volume = "86",
  pages = "094513",
  year = "2012",
}

@article{Leskovec:2012gb,
  author = "Leskovec, Luka and Prelovsek, Sasa",
  title = "{Scattering phase shifts for two particles of different mass and
           non-zero total momentum in lattice QCD}",
  eprint = "1202.2145",
  archivePrefix = "arXiv",
  primaryClass = "hep-lat",
  doi = "10.1103/PhysRevD.85.114507",
  journal = "Phys. Rev. D",
  volume = "85",
  pages = "114507",
  year = "2012",
}

@article{Fu:2011wc,
  author = "Fu, Ziwen",
  title = "{Lattice study on $\pi K $ scattering with moving wall source}",
  eprint = "1110.1422",
  archivePrefix = "arXiv",
  primaryClass = "hep-lat",
  doi = "10.1103/PhysRevD.85.074501",
  journal = "Phys. Rev. D",
  volume = "85",
  pages = "074501",
  year = "2012",
}

@article{Fu:2012tj,
  author = "Fu, Ziwen and Fu, Kan",
  title = "{Lattice QCD study on $K^\ast(892)$ meson decay width}",
  eprint = "1209.0350",
  archivePrefix = "arXiv",
  primaryClass = "hep-lat",
  doi = "10.1103/PhysRevD.86.094507",
  journal = "Phys. Rev. D",
  volume = "86",
  pages = "094507",
  year = "2012",
}

@article{Prelovsek:2013ela,
  author = "Prelovsek, Sasa and Leskovec, Luka and Lang, C. B. and Mohler,
            Daniel",
  title = "{K $\pi$ Scattering and the K* Decay width from Lattice QCD}",
  eprint = "1307.0736",
  archivePrefix = "arXiv",
  primaryClass = "hep-lat",
  reportNumber = "FERMILAB-PUB-13-251-T",
  doi = "10.1103/PhysRevD.88.054508",
  journal = "Phys. Rev. D",
  volume = "88",
  number = "5",
  pages = "054508",
  year = "2013",
}

@article{Brett:2018jqw,
  author = {Brett, Ruair\'\i{} and Bulava, John and Fallica, Jacob and Hanlon,
            Andrew and H\"orz, Ben and Morningstar, Colin},
  title = "{Determination of $s$- and $p$-wave $I=1/2$ $K\pi$ scattering
           amplitudes in $N_{\mathrm{f}}=2+1$ lattice QCD}",
  eprint = "1802.03100",
  archivePrefix = "arXiv",
  primaryClass = "hep-lat",
  reportNumber = "MITP-18-010, CP3-ORIGINS-2018-5",
  doi = "10.1016/j.nuclphysb.2018.05.008",
  journal = "Nucl. Phys. B",
  volume = "932",
  pages = "29--51",
  year = "2018",
}

@article{Rendon:2020rtw,
  author = "Rendon, Gumaro and Leskovec, Luka and Meinel, Stefan and Negele,
            John and Paul, Srijit and Petschlies, Marcus and Pochinsky, Andrew
            and Silvi, Giorgio and Syritsyn, Sergey",
  title = "{$I=1/2$ $S$-wave and $P$-wave $K\pi$ scattering and the $\kappa$ and
           $K^*$ resonances from lattice QCD}",
  eprint = "2006.14035",
  archivePrefix = "arXiv",
  primaryClass = "hep-lat",
  reportNumber = "JLAB-THY-20-3213",
  doi = "10.1103/PhysRevD.102.114520",
  journal = "Phys. Rev. D",
  volume = "102",
  number = "11",
  pages = "114520",
  year = "2020",
}

@article{Boyle:2024grr,
  author = {Boyle, Peter and Erben, Felix and G\"ulpers, Vera and Hansen,
            Maxwell T. and Joswig, Fabian and Marshall, Michael and Lachini,
            Nelson Pitanga and Portelli, Antonin},
  title = "{Physical-mass calculation of \ensuremath{\rho}(770) and K*(892)
           resonance parameters via \ensuremath{\pi}\ensuremath{\pi} and K
           \ensuremath{\pi} scattering amplitudes from lattice QCD}",
  eprint = "2406.19193",
  archivePrefix = "arXiv",
  primaryClass = "hep-lat",
  reportNumber = "CERN-TH-2024-087",
  doi = "10.1103/PhysRevD.111.054510",
  journal = "Phys. Rev. D",
  volume = "111",
  number = "5",
  pages = "054510",
  year = "2025",
}

@article{Dudek:2014qha,
  author = "Dudek, Jozef J. and Edwards, Robert G. and Thomas, Christopher E.
            and Wilson, David J.",
  collaboration = "Hadron Spectrum",
  title = "{Resonances in coupled $\pi K -\eta K$ scattering from quantum
           chromodynamics}",
  eprint = "1406.4158",
  archivePrefix = "arXiv",
  primaryClass = "hep-ph",
  reportNumber = "JLAB-THY-14-1891, DAMTP-2014-36",
  doi = "10.1103/PhysRevLett.113.182001",
  journal = "Phys. Rev. Lett.",
  volume = "113",
  number = "18",
  pages = "182001",
  year = "2014",
}

@article{Yan:2024yuq,
  author = "Yan, Haobo and Liu, Chuan and Liu, Liuming and Meng, Yu and Xing,
            Hanyang",
  title = "{Pion mass dependence in D\ensuremath{\pi} scattering and the
           D0*(2300) resonance from lattice QCD}",
  eprint = "2404.13479",
  archivePrefix = "arXiv",
  primaryClass = "hep-lat",
  doi = "10.1103/PhysRevD.111.014503",
  journal = "Phys. Rev. D",
  volume = "111",
  number = "1",
  pages = "014503",
  year = "2025",
}

@article{Yan:2025jlq,
    author = "Yan, Haobo and Liu, Chuan and Liu, Liuming and Meng, Yu",
    title = "{Construction of general N-body lattice operators with arbitrary momenta}",
    eprint = "2507.16070",
    archivePrefix = "arXiv",
    primaryClass = "hep-lat",
    doi = "10.1007/JHEP10(2025)210",
    journal = "JHEP",
    volume = "10",
    pages = "210",
    year = "2025"
}

@article{Cao:2025hqm,
    author = "Cao, Xiong-Hui and Guo, Feng-Kun and Guo, Zhi-Hui and Li, Qu-Zhi",
    title = "{Revisiting Roy-Steiner-equation analysis of pion-kaon scattering from lattice QCD data}",
    eprint = "2506.10619",
    archivePrefix = "arXiv",
    primaryClass = "hep-ph",
    doi = "10.1103/hn8j-95vn",
    journal = "Phys. Rev. D",
    volume = "112",
    number = "3",
    pages = "034042",
    year = "2025"
}

@article{Fu:2011xz,
  author = "Fu, Ziwen",
  title = "{Rummukainen-Gottlieb's formula on two-particle system with different
           mass}",
  eprint = "1110.0319",
  archivePrefix = "arXiv",
  primaryClass = "hep-lat",
  doi = "10.1103/PhysRevD.85.014506",
  journal = "Phys. Rev. D",
  volume = "85",
  pages = "014506",
  year = "2012",
}

@article{Zheng:2003rw,
  author = "Zheng, H. Q. and Zhou, Z. Y. and Qin, G. Y. and Xiao, Zhiguang and
            Wang, J. J. and Wu, N.",
  title = "{The kappa resonance in s wave pi K scatterings}",
  eprint = "hep-ph/0310293",
  archivePrefix = "arXiv",
  doi = "10.1016/j.nuclphysa.2003.12.021",
  journal = "Nucl. Phys. A",
  volume = "733",
  pages = "235--261",
  year = "2004",
}

@article{Guo:2007ff,
  author = "Guo, Z. H. and Sanz Cillero, J. J. and Zheng, H. Q.",
  title = "{Partial waves and large N(C) resonance sum rules}",
  eprint = "hep-ph/0701232",
  archivePrefix = "arXiv",
  doi = "10.1088/1126-6708/2007/06/030",
  journal = "JHEP",
  volume = "06",
  pages = "030",
  year = "2007",
}

@article{Guo:2007hm,
  author = "Guo, Z. H. and Sanz-Cillero, J. J. and Zheng, H. Q.",
  title = "{O(p**6) extension of the large - N(C) partial wave dispersion
           relations}",
  eprint = "0710.2163",
  archivePrefix = "arXiv",
  primaryClass = "hep-ph",
  doi = "10.1016/j.physletb.2008.01.073",
  journal = "Phys. Lett. B",
  volume = "661",
  pages = "342--347",
  year = "2008",
}

@article{Zhou:2004ms,
  author = "Zhou, Z. Y. and Qin, G. Y. and Zhang, P. and Xiao, Zhiguang and
            Zheng, H. Q. and Wu, N.",
  title = "{The Pole structure of the unitary, crossing symmetric low energy pi
           pi scattering amplitudes}",
  eprint = "hep-ph/0406271",
  archivePrefix = "arXiv",
  doi = "10.1088/1126-6708/2005/02/043",
  journal = "JHEP",
  volume = "02",
  pages = "043",
  year = "2005",
}

@article{Wang:2018nwi,
  author = "Wang, Yu-Fei and Yao, De-Liang and Zheng, Han-Qing",
  title = "{New insights on low energy $\pi N$ scattering amplitudes:
           comprehensive analyses at $\mathcal{O}(p^3)$ level}",
  eprint = "1811.09748",
  archivePrefix = "arXiv",
  primaryClass = "hep-ph",
  doi = "10.1088/1674-1137/43/6/064110",
  journal = "Chin. Phys. C",
  volume = "43",
  number = "6",
  pages = "064110",
  year = "2019",
}

@article{LiQuZhi:2021nrq,
  author = "Li, Qu-Zhi and Ma, Yao and  Niu, Wen-Qi and Wang, Yu-Fei and Zheng, Han-Qing 
            ",
  title = "{An N/D study of the S 11 channel {\ensuremath{\pi}}N scattering
           amplitude *}",
  eprint = "2102.00977",
  archivePrefix = "arXiv",
  primaryClass = "nucl-th",
  doi = "10.1088/1674-1137/ac31a4",
  journal = "Chin. Phys. C",
  volume = "46",
  number = "2",
  pages = "023104",
  year = "2022",
}

@article{Bali:2015gji,
  author = {Bali, Gunnar S. and Collins, Sara and Cox, Antonio and Donald,
            Gordon and G{\"o}ckeler, Meinulf and Lang, C. B. and Sch{\"a}fer,
            Andreas},
  collaboration = "RQCD",
  title = "{$\rho$ and $K^*$ resonances on the lattice at nearly physical quark
           masses and $N_f=2$}",
  eprint = "1512.08678",
  archivePrefix = "arXiv",
  primaryClass = "hep-lat",
  doi = "10.1103/PhysRevD.93.054509",
  journal = "Phys. Rev. D",
  volume = "93",
  number = "5",
  pages = "054509",
  year = "2016",
}

@article{Nebreda:2010wv,
  author = "Nebreda, J. and Pelaez., J. R.",
  title = "{Strange and non-strange quark mass dependence of elastic light
           resonances from SU(3) Unitarized Chiral Perturbation Theory to one
           loop}",
  eprint = "1001.5237",
  archivePrefix = "arXiv",
  primaryClass = "hep-ph",
  doi = "10.1103/PhysRevD.81.054035",
  journal = "Phys. Rev. D",
  volume = "81",
  pages = "054035",
  year = "2010",
}

@article{Pelaez:2020gnd,
  author = "Pel{\'a}ez, Jos{\'e} R. and Rodas, Arkaitz",
  title = "{Dispersive {\ensuremath{\pi}}K{\textrightarrow}{\ensuremath{\pi}}K
           and {\ensuremath{\pi}}{\ensuremath{\pi}}{\textrightarrow}KK̄
           amplitudes from scattering data, threshold parameters, and the
           lightest strange resonance {\ensuremath{\kappa}} or K0{\ensuremath{*}}
           (700)}",
  eprint = "2010.11222",
  archivePrefix = "arXiv",
  primaryClass = "hep-ph",
  reportNumber = "JLAB-THY-20-3276",
  doi = "10.1016/j.physrep.2022.03.004",
  journal = "Phys. Rept.",
  volume = "969",
  pages = "1--126",
  year = "2022",
}

@article{Lang:2012sv,
  author = "Lang, C. B. and Leskovec, Luka and Mohler, Daniel and Prelovsek,
            Sasa",
  title = "{K pi scattering for isospin 1/2 and 3/2 in lattice QCD}",
  eprint = "1207.3204",
  archivePrefix = "arXiv",
  primaryClass = "hep-lat",
  doi = "10.1103/PhysRevD.86.054508",
  journal = "Phys. Rev. D",
  volume = "86",
  pages = "054508",
  year = "2012",
}

@article{Zhou:2006wm,
  author = "Zhou, Z. Y. and Zheng, H. Q.",
  title = "{An improved study of the kappa resonance and the non-exotic $s$ wave
           $\pi K$ scatterings up to $\sqrt{s}=2.1$GeV of LASS data}",
  eprint = "hep-ph/0603062",
  archivePrefix = "arXiv",
  doi = "10.1016/j.nuclphysa.2006.06.170",
  journal = "Nucl. Phys. A",
  volume = "775",
  pages = "212--223",
  year = "2006",
}

@article{Pelaez:2020uiw,
  author = "Pel{\'a}ez, J. R. and Rodas, A.",
  title = "{Determination of the lightest strange resonance $K_0^*(700)$ or $
           \kappa$, from a dispersive data analysis}",
  eprint = "2001.08153",
  archivePrefix = "arXiv",
  primaryClass = "hep-ph",
  reportNumber = "JLAB-THY-20-3166",
  doi = "10.1103/PhysRevLett.124.172001",
  journal = "Phys. Rev. Lett.",
  volume = "124",
  number = "17",
  pages = "172001",
  year = "2020",
}

@article{Miao:2004gy,
  author = "Miao, Chuan and Du, Xin-ing and Meng, Guang-wei and Liu, Chuan",
  title = "{Lattice study on kaon pion scattering length in the I = 3/2 channel}
           ",
  eprint = "hep-lat/0403028",
  archivePrefix = "arXiv",
  doi = "10.1016/j.physletb.2004.05.073",
  journal = "Phys. Lett. B",
  volume = "595",
  pages = "400--407",
  year = "2004",
}

@article{Nagata:2008wk,
  author = "Nagata, Junichi and Muroya, Shin and Nakamura, Atsushi",
  title = "{Lattice study of K pi scattering in I = 3/2 and 1/2}",
  eprint = "0812.1753",
  archivePrefix = "arXiv",
  primaryClass = "hep-lat",
  doi = "10.1103/PhysRevC.84.019904",
  journal = "Phys. Rev. C",
  volume = "80",
  pages = "045203",
  year = "2009",
  note = "[Erratum: Phys.Rev.C 84, 019904 (2011)]",
}

@article{Sasaki:2013vxa,
  author = "Sasaki, Kiyoshi and Ishizuka, Naruhito and Oka, Makoto and Yamazaki,
            Takeshi",
  collaboration = "PACS-CS",
  title = "{Scattering lengths for two pseudoscalar meson systems}",
  eprint = "1311.7226",
  archivePrefix = "arXiv",
  primaryClass = "hep-lat",
  doi = "10.1103/PhysRevD.89.054502",
  journal = "Phys. Rev. D",
  volume = "89",
  number = "5",
  pages = "054502",
  year = "2014",
  note = "[Erratum: Phys.Rev.D 105, 019901 (2022)]",
}

@article{Prelovsek:2010kg,
  author = "Prelovsek, Sasa and Draper, Terrence and Lang, Christian B. and
            Limmer, Markus and Liu, Keh-Fei and Mathur, Nilmani and Mohler,
            Daniel",
  title = "{Lattice study of light scalar tetraquarks with I=0,2,1/2,3/2: Are {
           \textbackslash}sigma and {\textbackslash}kappa tetraquarks?}",
  eprint = "1005.0948",
  archivePrefix = "arXiv",
  primaryClass = "hep-lat",
  doi = "10.1103/PhysRevD.82.094507",
  journal = "Phys. Rev. D",
  volume = "82",
  pages = "094507",
  year = "2010",
}

@article{Yan:2025mdm,
  author = {Yan, Haobo and Mai, Maxim and Garofalo, Marco and Feng, Yuchuan and
            D{\"o}ring, Michael and Liu, Chuan and Liu, Liuming and Mei{\ss}ner,
            Ulf-G. and Urbach, Carsten},
  title = "{Emergence of the $\pi(1300)$ Resonance from Lattice QCD}",
  eprint = "2510.09476",
  archivePrefix = "arXiv",
  primaryClass = "hep-lat",
  month = "10",
  year = "2025",
}

@article{Guo:2013nja,
  author = "Guo, Feng-Kun and Liu, Liuming and Meissner, Ulf-G. and Wang, Ping",
  title = "{Tetraquarks, hadronic molecules, meson-meson scattering and
           disconnected contributions in lattice QCD}",
  eprint = "1308.2545",
  archivePrefix = "arXiv",
  primaryClass = "hep-lat",
  doi = "10.1103/PhysRevD.88.074506",
  journal = "Phys. Rev. D",
  volume = "88",
  pages = "074506",
  year = "2013",
}

@article{Alexandrou:2012rm,
  author = "Alexandrou, Constantia and Daldrop, Jan Oliver and Dalla Brida,
            Mattia and Gravina, Mario and Scorzato, Luigi and Urbach, Carsten and
            Wagner, Marc",
  title = "{Lattice investigation of the scalar mesons $a_0$(980) and $\kappa$
           using four-quark operators}",
  eprint = "1212.1418",
  archivePrefix = "arXiv",
  primaryClass = "hep-lat",
  doi = "10.1007/JHEP04(2013)137",
  journal = "JHEP",
  volume = "04",
  pages = "137",
  year = "2013",
}

@article{Yan:2024gwp,                                                           
    author = "Yan, Haobo and Mai, Maxim and Garofalo, Marco and Mei{\ss}ner,    
Ulf-G. and Liu, Chuan and Liu, Liuming and Urbach, Carsten",                    
    title = "{{\ensuremath{\omega}} Meson from Lattice QCD}",                   
    eprint = "2407.16659",                                                      
    archivePrefix = "arXiv",                                                    
    primaryClass = "hep-lat",                                                   
    doi = "10.1103/PhysRevLett.133.211906",                                     
    journal = "Phys. Rev. Lett.",                                               
    volume = "133",                                                             
    number = "21",                                                              
    pages = "211906",                                                           
    year = "2024"                                                               
}

@article{CLQCD:2026jyw,
    author = "Lin, Tong-Wei and others",
    collaboration = "CLQCD",
    title = "{Impact of dynamical charm quark and mixed action effect on light hadron masses and decay constants}",
    eprint = "2603.04230",
    archivePrefix = "arXiv",
    primaryClass = "hep-lat",
    doi = "10.1088/1572-9494/ae6cc4",
    journal = "Commun. Theor. Phys.",
    volume = "78",
    number = "9",
    pages = "095201",
    year = "2026"
}

@article{Yi:2025bnh,                                                            
    author = "Yi, Jing-Yu and Liang, Ze-Rui and Liu, Liuming and Yao, De-Liang",
    title = "{Low-energy interactions between doubly charmed baryons and        
Goldstone bosons from lattice QCD}",                                            
    eprint = "2511.12611",                                                      
    archivePrefix = "arXiv",                                                    
    primaryClass = "hep-lat",                                                   
    month = "11",                                                               
    year = "2025"                                                               
}

@article{Shi:2025ogt,                                                           
    author = "Shi, Pan-Pan and Guo, Feng-Kun and Liu, Chuan and Liu, Liuming and
Sun, Peng and Wu, Jia-Jun and Xing, Hanyang",                                   
    title = "{Low-energy $DD$ scattering in lattice QCD}",                      
    eprint = "2502.07438",                                                      
    archivePrefix = "arXiv",                                                    
    primaryClass = "hep-lat",                                                   
    month = "2",                                                                
    year = "2025"                                                               
}

@article{Wang:2025hew,                                                          
    author = "Wang, Zhengli and Leinweber, Derek B. and Liu, Chuan and Liu,     
Liuming and Sun, Peng and Thomas, Anthony W. and Wu, Jia-jun and Xing, Hanyang  
and Yu, Kang",                                                                  
    collaboration = "CLQCD",                                                    
    title = "{Spectral parameters of the {\ensuremath{\rho}} resonance from     
lattice QCD}",                                                                  
    eprint = "2502.03700",                                                      
    archivePrefix = "arXiv",                                                    
    primaryClass = "hep-lat",                                                   
    reportNumber = "ADP-25-4/1266",                                             
    doi = "10.1007/JHEP08(2025)064",                                            
    journal = "JHEP",                                                           
    volume = "08",                                                              
    pages = "064",                                                              
    year = "2025"                                                               
}

@article{Xing:2025uai,                                                          
    author = "Xing, Hanyang and Geng, Yiqi and Liu, Chuan and Liu, Liuming and  
Sun, Peng and Wu, Jiajun and Yan, Zhicheng and Zhu, Ruilin",                    
    title = "{Lattice QCD study of                                              
{\ensuremath{\Lambda}}$_{c}${\ensuremath{\Lambda}}$_{c}$ scattering*}",         
    eprint = "2502.05546",                                                      
    archivePrefix = "arXiv",                                                    
    primaryClass = "hep-lat",                                                   
    doi = "10.1088/1674-1137/adc11f",                                           
    journal = "Chin. Phys. C",                                                  
    volume = "49",                                                              
    number = "6",                                                               
    pages = "063107",                                                           
    year = "2025"                                                               
}

@article{Kim:2005gf,
    author = "Kim, C. h. and Sachrajda, C. T. and Sharpe, Stephen R.",
    title = "{Finite-volume effects for two-hadron states in moving frames}",
    eprint = "hep-lat/0507006",
    archivePrefix = "arXiv",
    reportNumber = "UW-PT-05-16, SHEP-0518",
    doi = "10.1016/j.nuclphysb.2005.08.029",
    journal = "Nucl. Phys. B",
    volume = "727",
    pages = "218--243",
    year = "2005"
}

@article{He:2005ey,
    author = "He, Song and Feng, Xu and Liu, Chuan",
    title = "{Two particle states and the S-matrix elements in multi-channel scattering}",
    eprint = "hep-lat/0504019",
    archivePrefix = "arXiv",
    doi = "10.1088/1126-6708/2005/07/011",
    journal = "JHEP",
    volume = "07",
    pages = "011",
    year = "2005"
}

@article{Briceno:2014oea,
    author = "Briceno, Raul A.",
    title = "{Two-particle multichannel systems in a finite volume with arbitrary spin}",
    eprint = "1401.3312",
    archivePrefix = "arXiv",
    primaryClass = "hep-lat",
    reportNumber = "JLAB-THY-14-1833",
    doi = "10.1103/PhysRevD.89.074507",
    journal = "Phys. Rev. D",
    volume = "89",
    number = "7",
    pages = "074507",
    year = "2014"
}

@article{Luscher:1986pf,
    author = "Luscher, M.",
    title = "{Volume Dependence of the Energy Spectrum in Massive Quantum Field Theories. 2. Scattering States}",
    reportNumber = "DESY-86-034",
    doi = "10.1007/BF01211097",
    journal = "Commun. Math. Phys.",
    volume = "105",
    pages = "153--188",
    year = "1986"
}

@article{He:2002ut,
    author = "He, Jianyang and Xiao, Zhiguang and Zheng, H. Q.",
    title = "{The Constraints of unitary on $\pi\pi$ scattering dispersion relations}",
    eprint = "hep-ph/0201257",
    archivePrefix = "arXiv",
    doi = "10.1016/S0370-2693(02)01811-7",
    journal = "Phys. Lett. B",
    volume = "536",
    pages = "59--66",
    year = "2002",
    note = "[Erratum: Phys.Lett.B 549, 362--363 (2002)]"
}

@article{Descotes-Genon:2006sdr,
    author = "Descotes-Genon, S. and Moussallam, B.",
    title = "{The K*0 (800) scalar resonance from Roy-Steiner representations of pi K scattering}",
    eprint = "hep-ph/0607133",
    archivePrefix = "arXiv",
    reportNumber = "LPT-ORSAY-06-43",
    doi = "10.1140/epjc/s10052-006-0036-2",
    journal = "Eur. Phys. J. C",
    volume = "48",
    pages = "553",
    year = "2006"
}

@article{Fu:2026ntz,
    author = "Fu, Ziwen and Li, Qu-Zhi and Wang, Jun",
    title = "{I=32 {\ensuremath{\pi}}K s-wave scattering length from lattice QCD}",
    eprint = "2601.01205",
    archivePrefix = "arXiv",
    primaryClass = "hep-lat",
    doi = "10.1103/c66h-96wh",
    journal = "Phys. Rev. D",
    volume = "113",
    number = "3",
    pages = "034501",
    year = "2026"
}
\end{document}